\documentclass[iop]{emurateapj}

\usepackage{fancybox}
\usepackage{ulem}
\usepackage{ascmac}
\usepackage{color}
\usepackage{bm}

\newcommand{\Af}{A\_N4096L400}
\newcommand{\As}{A\_N4096L200}
\newcommand{\Bf}{B\_N2048L200}
\newcommand{\Mf}{EQ\ref{eq:doublepower}\_c1p35}
\newcommand{\Ms}{EQ\ref{eq:doublepower}\_cNFW}
\newcommand{\Mt}{NFW\_cNFW}

\shorttitle{Hierarchical Formation of Dark Matter Halos and the Free Streaming Scale}
\shortauthors{Ishiyama}

\begin{document}

\title{Hierarchical Formation of Dark Matter Halos and the Free Streaming Scale}

\author{ 
TOMOAKI \textsc{Ishiyama}\altaffilmark{1}
} 

\altaffiltext{1}{Center for Computational Science, University of Tsukuba, 1-1-1, 
Tennodai, Tsukuba, Ibaraki, 305-8577, Japan ; ishiyama@ccs.tsukuba.ac.jp}

\begin{abstract} 
The smallest dark matter halos are formed first in the early
universe.  According to recent studies, the central density cusp is much 
steeper in these halos than in larger halos and scales as
$\rho \propto r^{-(1.5\mbox{\scriptsize --}1.3)}$.
We present results of very large cosmological $N$-body 
simulations of the hierarchical formation and evolution of halos 
over a wide mass range, 
beginning from the formation of the smallest halos. We confirmed early 
studies that the inner density cusps are steeper in halos at the free 
streaming scale. The cusp slope gradually becomes shallower as the halo mass 
increases. 
The slope of halos 50 times more massive than the smallest 
halo is approximately $-1.3$. 
No strong correlation exists between inner slope and the collapse epoch.
The cusp slope of halos above the free streaming scale seems to be 
reduced primarily due to major merger processes.
The concentration, estimated at the present 
universe, is predicted to be $60\mbox{--}70$, 
consistent with theoretical models and earlier simulations, 
and ruling out simple power law mass--concentration relations.
Microhalos could still exist in the present universe with the same steep density 
profiles.
\end{abstract}


%%%%%%%%%%%%%%%%%%%%%%%%%%%%%%%%%%%%%%%%%%%%%%%%%%%%%%%%%%%%%%%%%%%%%%%%%%%%%%
%%%%%%%%%%%%%%%%%%%%%%%%%%%%%%%%%%%%%%%%%%%%%%%%%%%%%%%%%%%%%%%%%%%%%%%%%%%%%%
%%%%%%%%%%%%%%%%%%%%%%%%%%%%%%%%%%%%%%%%%%%%%%%%%%%%%%%%%%%%%%%%%%%%%%%%%%%%%%
%%%%%%%%%%%%%%%%%%%%%%%%%%%%%%%%%%%%%%%%%%%%%%%%%%%%%%%%%%%%%%%%%%%%%%%%%%%%%%
\section{Introduction}\label{sec:intro}

According to a concordance cold dark matter (CDM) theory, dark matter halos
hierarchically evolve in an expanding universe.  The smallest dark
matter microhalos first undergo gravitational collapse, and then merge
to form larger halos.  The size of the smallest halo is determined by
the free streaming scale of the dark matter particles.  If dark matter
comprises the lightest supersymmetric particle (the neutralino of mass
approximately 100 GeV), a cutoff in the matter power spectrum of the
very early universe should result from free streaming damping of this
particle.  The estimated corresponding mass of the smallest microhalos
is approximately earth-mass, $3.5 \times 10^{-9}\mbox{\scriptsize --}8.4 \times
10^{-6} M_{\odot}$ (
\citealt{Zybin1999, Hofmann2001, Berezinsky2003, Green2004, Loeb2005,
Bertschinger2006, Profumo2006, Berezinsky2008}). 
However, this interval could be increased by uncertainty in supersymmetry theory.

If many earth-mass microhalos exist at present universe, they could
significantly enhance gamma-ray signals by neutralino
self-annihilation \citep{Berezinsky2003, Diemand2005, Berezinsky2008,
  Ishiyama2010}.  Many studies have debated whether gamma-rays
generated by neutralino annihilation in subhalos and microhalos are
observable in indirect dark matter detection experiments
\citep[e.g.,][]{Berezinsky2003, Koushiappas2004, Oda2005,
  Colafrancesco2006, 
  Koushiappas2006, Goerdt2007, Diemand2007,
  Ando2008, Diemand2008, Kuhlen2008,
  Springel2008b, Lee2009, Giocoli2009, Ishiyama2010, Kamionkowski2010,
  Schneider2010, Charbonnier2011, Pieri2011, Belotsky2012,
  Blanchet2012, Mack2013, Ng2013}.  Since the gamma-ray flux is
proportional to the square of the local dark matter density, the
annihilation signal from the Milky Way halo should largely depend on
the halo's fine structure.

Analytic studies and cosmological simulations have predicted that the
subhalo mass function scales as 
$dn/dm \propto m^{-(2\mbox{\scriptsize --}1.8)}$ 
\citep[e.g.,][]{Berezinsky2003, Ishiyama2009}, although no
consensus has yet emerged.  Thus, smaller subhalos are abundant in the
Milky Way, and some of them will probably pass near the Sun.  The
survivability of microhalos and larger subhalos in the Milky Way
depends on their structure. This suggests that the structure of the
smallest microhalos can determine the fine structure of the dark matter
halo.  Therefore, to evaluate the observable
annihilation flux, we must understand the density structures of
microhalos.

Historically, the structures of relatively large halos (galaxy-sized
to cluster-sized) have been investigated by cosmological $N$-body
simulations \citep[e.g.,][]{Springel2008,Diemand2008,Ishiyama2013}.  In
most works, the structures of these halos follow the
Navarro--Frenk--White (NFW) profile \citep{Navarro1996} or the Einasto
profile \citep{Einasto1965}.  In these profiles, the slopes of radial
density profiles approximate $-1$ in the inner region, gradually
increasing $-3$ toward the outer region.  Some of the above cited
studies have estimated the gamma-ray flux from microhalos assuming
that microhalos follow these profiles.
However, unlike massive halos contain many subhalos, microhalos
contain no subhalos by definition and do not form by hierarchical
merging of smaller halos. Thus, the structure of microhalos may largely
differ from that of larger halos. 

The structure of microhalos was first investigated in numerical
simulations performed by \citet{Diemand2005}.  They successfully
fitted the density profiles of microhalos to the form, $\rho(r)
\propto r^{-\gamma}$, where the slope $\gamma$ ranges from 1.5 to 2,
down to radii as low as $\sim 10^{-3}$ pc.  However, the mass
resolution in the simulation of \citet{Diemand2005} was $1.2 \times
10^{-10} M_{\odot}$, insufficient to resolve the central structures of
microhalos.

\citet{Ishiyama2010} improved the mass and spatial resolutions in the
earlier study by 100 and 20 times, respectively. After numerous
simulations were performed at the higher resolution, they found that the
central density cusps of microhalos scale much more steeply than those
of larger halos, with $\rho \propto r^{-1.5}$. These results are
supported by recent similar simulations \citep{Anderhalden2013}, in
which inner profiles scale as 
$\rho \propto r^{-(1.4\mbox{\scriptsize --}1.3)}$.

Such steep cusps in the density profile may largely 
impact on indirect experimental searches for dark matter. 
The contribution of  microhalos, based on the density
profiles derived from cosmological simulations, are estimated in
\citet{Diemand2005}, \citet{Ishiyama2010}, and \citet{Anderhalden2013}.
However, their estimation relies on results of only a few simulated halos.
Statistical study, such as the distribution of
microhalo density profiles, requires a more extensive dataset.
\citet{Ishiyama2010} analytically estimated that the formation epoch
of microhalos is affected by larger scale density fluctuations.  They
predicted that the formation epoch of their simulated microhalos was
later than the average value.  Since halo concentration reflects the
cosmic density at which halos collapse, this suggests that their
microhalos were less concentrated than indicated by the average.
Therefore, to precisely predict the gamma-ray flux, the distribution of  
both the microhalo density profile shapes and of microhalo concentrations
must be elucidated. 

The statistics are
most effectively improved by simulating larger simulation boxes with
uniform mass resolution and unbiasedly analyzing all halos.  Although
we can now simulate halos with ultra-high resolution by re-simulation
method, selection bias is difficult to eliminate, and impedes the
acquisition of good statistical samples of microhalos.

Another difficulty is resolving the free streaming
scale with sufficient resolution.  
Two-body relaxation introduces numerical artifacts that
significantly reduce the central density of
microhalos.  Artificial fragmentation can occur at filaments below the
free streaming scale, as seen in warm dark matter simulations
\citep{Wang2007, Schneider2013, Angulo2013}.  To avoid such
numerical artifacts and obtain good statistics of well-resolved halos,
substantially many particles are required.  Such large simulations
are numerically challenging and consume huge computational resources.

Despite these difficulties, such simulations are worthy of attempt,
since they provide microhalo statistics and yield valuable structural
information on halos, whose masses exceed earth-mass by a few orders
of magnitude.  These halos are interesting targets since some of them
are formed by mergers of microhalos.  These halos could be
structurally different from larger halos, and could resemble the
smallest microhalos.  Along with microhalos, these small halos may
significantly contribute to the gamma-ray signal.  The structural
statistics of these halos may help to elucidate the mass scale at
which the density profile transits to steeper cusps.  
\citet{Diemand2006} performed high-resolution simulation for a $0.014M_\odot$ halo and analyzed it. 
However, there is only one halo, 
statistical study over a wide mass range is needed. 

Another interesting point is that our simulations are closely related
to warm dark matter simulations.  Our simulations would give a huge
opportunity to understand warm dark matter halos, since the cutoff in
the matter power spectrum is imposed by a similar mechanism, although
mass scales are largely different.

We address these questions by large and high resolution cosmological
$N$-body simulations.  We present the first study of halo structure
near the free streaming scale over a wide mass range. To reveal the
statistics of these halos in the early universe, we simulated a large
number of dark matter particles in sufficient volumes to reliably
sample these halos. \S \ref{sec:method} describes our simulation
method and its setup.  The structures of halos near the free streaming
scale, density profiles, profile distributions, dependence on the halo
formation epoch, and halo evolution are presented in \S
\ref{sec:result}. The contributions of these halos to 
gamma-ray signals by neutralino self-annihilation is discussed in \S
\ref{sec:discussion}.  The results are summarized in \S
\ref{sec:summary}.

%%%%%%%%%%%%%%%%%%%%%%%%%%%%%%%%%%%%%%%%%%%%%%%%%%%%%%%%%%%%%%%%%%%%%%%%%%%%%%
%%%%%%%%%%%%%%%%%%%%%%%%%%%%%%%%%%%%%%%%%%%%%%%%%%%%%%%%%%%%%%%%%%%%%%%%%%%%%%
%%%%%%%%%%%%%%%%%%%%%%%%%%%%%%%%%%%%%%%%%%%%%%%%%%%%%%%%%%%%%%%%%%%%%%%%%%%%%%
%%%%%%%%%%%%%%%%%%%%%%%%%%%%%%%%%%%%%%%%%%%%%%%%%%%%%%%%%%%%%%%%%%%%%%%%%%%%%%
\section{Initial Conditions and Numerical Method}\label{sec:method}

\begin{figure*}
\centering 
\includegraphics[width=18cm]{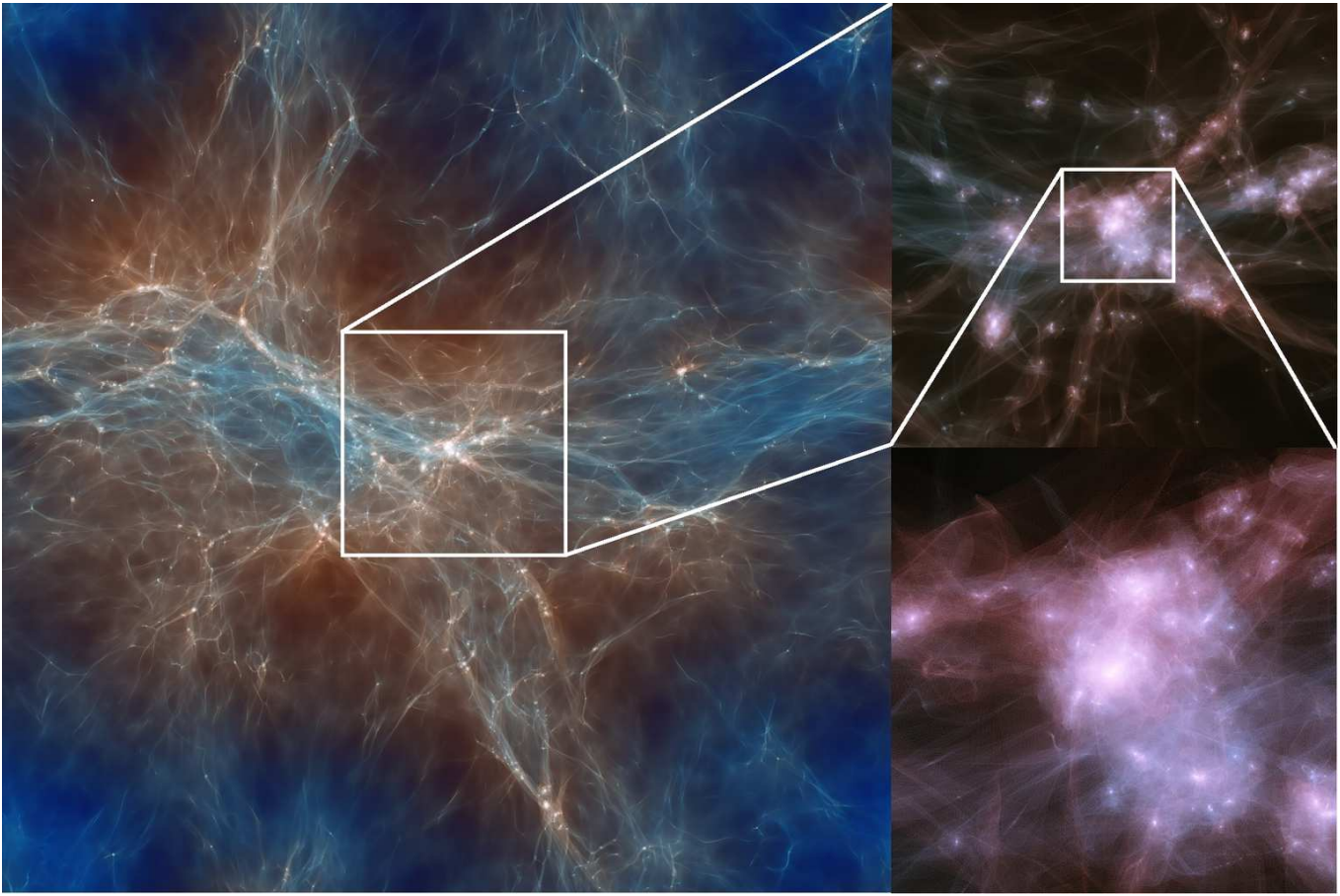} 
\caption{ 
Distribution of dark matter in the \As\ simulation at $z=32$, 
centered on the largest halo. 
Left panel shows the entire simulation volume. 
Right two panels are enlargements of the largest halo. 
The widths of the left, higher right, and lower right images 
correspond to 400, 100, and 25 comoving pc, respectively. 
}
\label{fig:snapshot}
\end{figure*}

We performed three large cosmological $N$-body simulations. The
cosmological parameters, namely, $\Omega_0=0.27$, $\lambda_0=0.73$, $h=0.7$,
and $\sigma_8=0.8$,  were based on the concordance
$\Lambda$CDM model \citep{Komatsu2011}.  Two different initial matter
power spectra were used.  
In two of the simulations, the power spectrum 
included the sharp cutoff imposed by the free streaming damping of dark
matter particles with a mass of 100GeV \citep{Green2004}.  
The third simulation ignored the effect of free streaming
damping.  We denote the former two simulations as model A, and the latter as model B.
The initial conditions were generated by a first-order Zeldovich
approximation at $z=400$.

In the model A simulations (with the cutoff imposed), the motions of
$4096^3$ particles in comoving boxes of side lengths 400 pc and 200 pc
were followed (these simulations are denoted \Af\ and \As,
respectively).  The particle masses were $3.4 \times 10^{-11}
M_{\odot}$ and $4.3 \times 10^{-12} M_{\odot}$, ensuring that halos at
the free streaming scale were represented by $\sim30,000$ and 
$\sim230,000$ particles, respectively.  Such high resolution protects the
halos from the artificial fragmentation as seen in warm dark
matter simulations \citep{Wang2007, Schneider2013, Angulo2013}.  The
respective gravitational Plummer softening lengths were $2.0 \times
10^{-4}$ pc and $1.0 \times 10^{-4}$ pc.  The model B simulation (with no
cutoff) followed the motions of $2048^3$ particles in comoving boxes
of side length 200 pc (this simulation is denoted \Bf).  The particle
mass was $3.4 \times 10^{-11} M_{\odot}$, and the gravitational
Plummer softening length was $2.0 \times 10^{-4}$ pc.  Simulations
\Af\ and \Bf\ were performed at the same mass and spatial resolution,
since both the number of particles and the simulated volume of
simulation \Bf\ were 1/8 those of simulation \Af.  To avoid unphysical
density fluctuations at small scales \citep{Colin2008}, thermal
velocities was not imposed in the initial conditions. 
The setup of the three simulations is detailed in Table \ref{tab1}.

\begin{table*}[t]
\centering
\caption
{Details of simulations.  Here, $N$, $L$, $\varepsilon$,
  $M$, and $m_{\rm DM}$ are the total number of particles, box
  length, softening length, mass resolution, and mass of
  dark matter particles, respectively  }\label{tab1}
\begin{tabular}{lccccc}
\hline\hline
Name  & $N$ & $L(\rm pc)$ & $\varepsilon(\rm pc)$ & $m (M_{\odot})$ & $m_{\rm DM}$ (GeV)\\
\hline
\Af & $4096^3$ & 400.0 & $2.0 \times 10^{-4}$ & $3.4 \times 10^{-11}$ & 100\\
\As & $4096^3$ & 200.0 & $1.0 \times 10^{-4}$ & $4.3 \times 10^{-12}$ & 100\\
\Bf & $2048^3$ & 200.0 & $2.0 \times 10^{-4}$ & $3.4 \times 10^{-11}$ & w/o cutoff\\
\hline 
\end{tabular}
\end{table*}

Simulations were performed by massively parallel TreePM code,
GreeM \citep{Ishiyama2009b, Ishiyama2012} 
\footnote{http://www.ccs.tsukuba.ac.jp/Astro/Members/ishiyama/greem}
on Aterui supercomputer at Center for
Computational Astrophysics, CfCA, of National Astronomical Observatory
of Japan, and the K computer at the RIKEN Advanced Institute for
Computational Science.  
The calculation of the tree force
was accelerated by the Phantom-GRAPE library \footnote{http://code.google.com/p/phantom-grape/}
\citep{Nitadori2006, Tanikawa2012, Tanikawa2013}
with support for AVX instruction set extension to the x86 architecture 
and the HPC-ACE architecture of the K computer.
PM calculations of $4096^3 (2048^3)$ particles simulations were performed on 
$2048^3 (1024^3)$ grid points
and the opening angle for the tree method was 0.5. Simulations were
terminated at $z=32$, where long-wavelength perturbations comparable 
to the box size were no longer negligible. 

Halos were identified by the spherical overdensity 
method\citep{Lacey1994}.  The virial radius of a halo $r_{\rm vir}$ is 
defined as the radius in which the spherical overdensity is $\Delta(z)= 
18\pi^2+ 82x - 39x^2$ times the critical value, where $x\equiv 
\Omega(z)-1$ \citep{Bryan1998}.  The virial mass of a halo $M_{\rm vir}$ 
is defined as the mass within the virial radius.  The most massive halos 
identified in \Af, \As, and \Bf\ simulations contained 170918717, 
48316099, and 10505232 particles, respectively.  The corresponding 
masses of these halos were $5.84 \times 10^{-3} M_{\odot}$, $2.08 \times 
10^{-4}M_{\odot}$, and $3.59 \times 10^{-4}M_{\odot}$.

Figure \ref{fig:snapshot} shows the dark matter distribution in the
\As\ simulation at $z=32$, centered on the largest halo.  Several
caustics are generated by nonlinear growth of long-wavelength
perturbations.  The distributions differ from those of large scale
structure simulations, and are analogous to those obtained in warm
dark matter simulations \citep[e.g.,][]{Bode2001}.

%%%%%%%%%%%%%%%%%%%%%%%%%%%%%%%%%%%%%%%%%%%%%%%%%%%%%%%%%%%%%%%%%%%%%%%%%%%%%%
%%%%%%%%%%%%%%%%%%%%%%%%%%%%%%%%%%%%%%%%%%%%%%%%%%%%%%%%%%%%%%%%%%%%%%%%%%%%%%
%%%%%%%%%%%%%%%%%%%%%%%%%%%%%%%%%%%%%%%%%%%%%%%%%%%%%%%%%%%%%%%%%%%%%%%%%%%%%%
%%%%%%%%%%%%%%%%%%%%%%%%%%%%%%%%%%%%%%%%%%%%%%%%%%%%%%%%%%%%%%%%%%%%%%%%%%%%%%
\section{Results}\label{sec:result}

\subsection{Density Profiles}\label{sec:profile}

\begin{figure}
\centering 
\includegraphics[width=8cm]{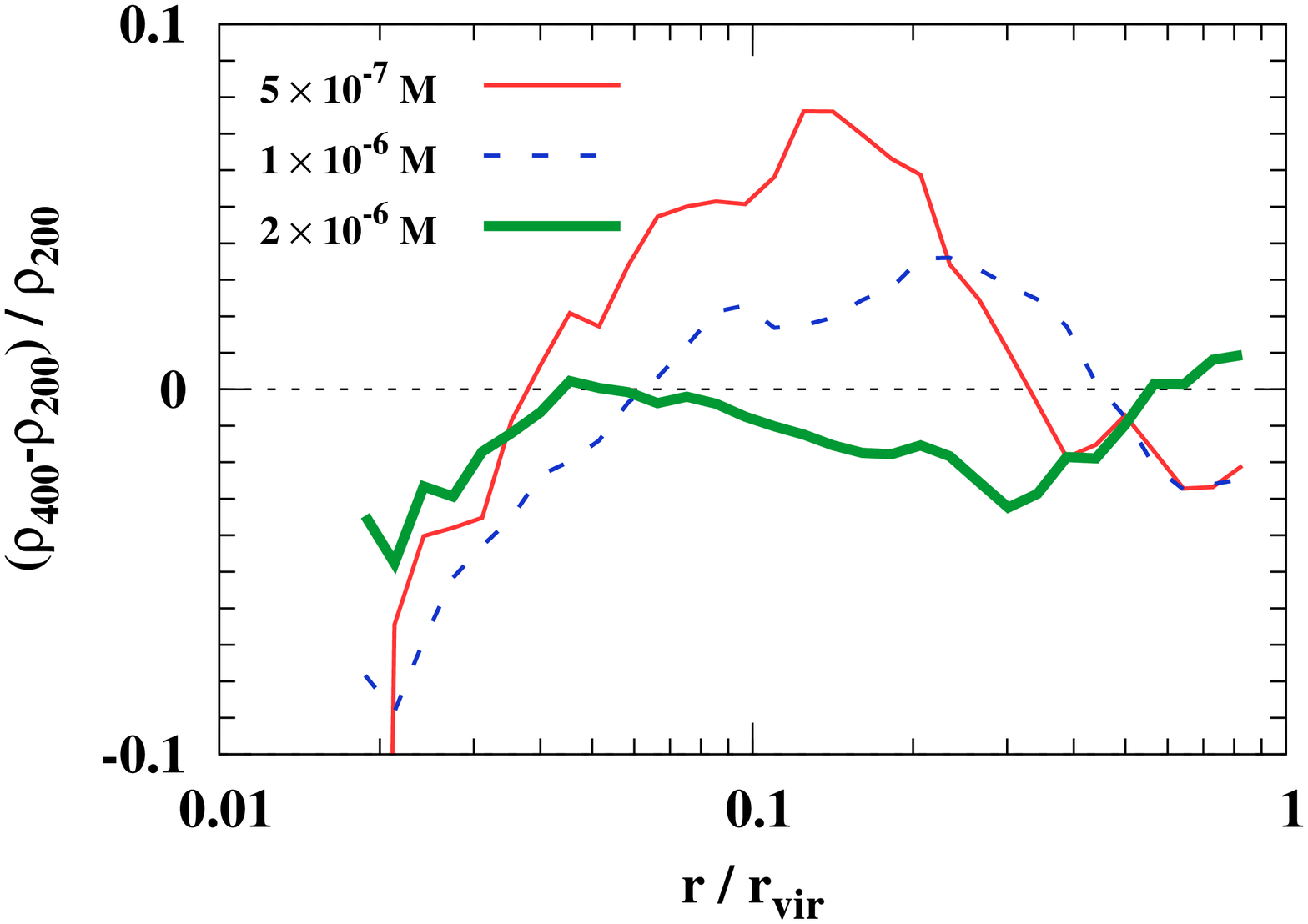} 
\caption{
Residuals of stacked radial density profiles
from simulation \Af\ ($\rho_{400}$) to \As\ ($\rho_{200}$)
as a function of radius (normalized by the virial radius).
The mass resolution in the \As\ simulation is eight times higher
than that of \Af.
The results of three different mass bins are presented. 
}
\label{fig:prof_conv}
\end{figure}

\begin{figure*}
\centering 
\includegraphics[width=5.3cm]{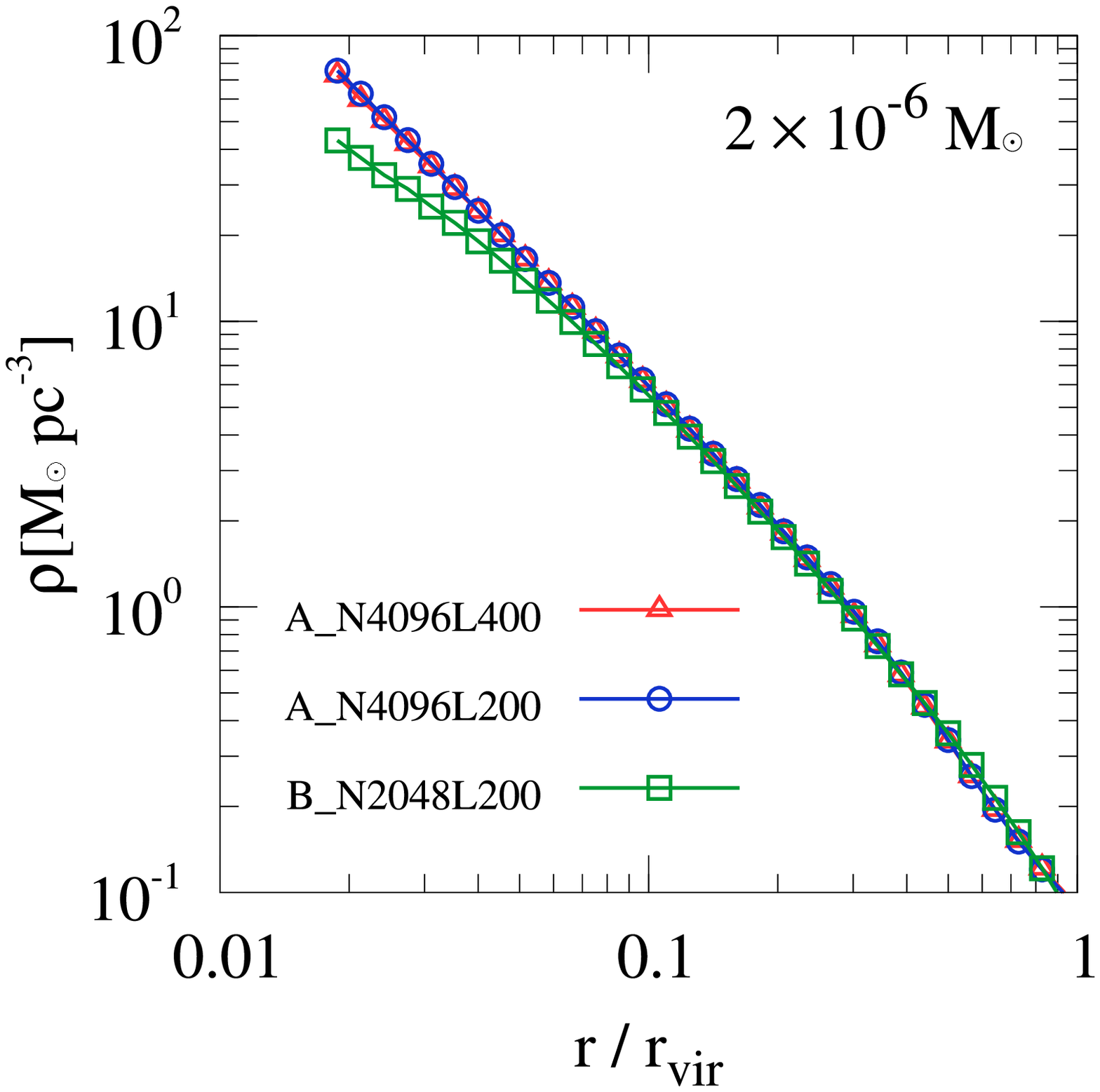} 
\includegraphics[width=5.3cm]{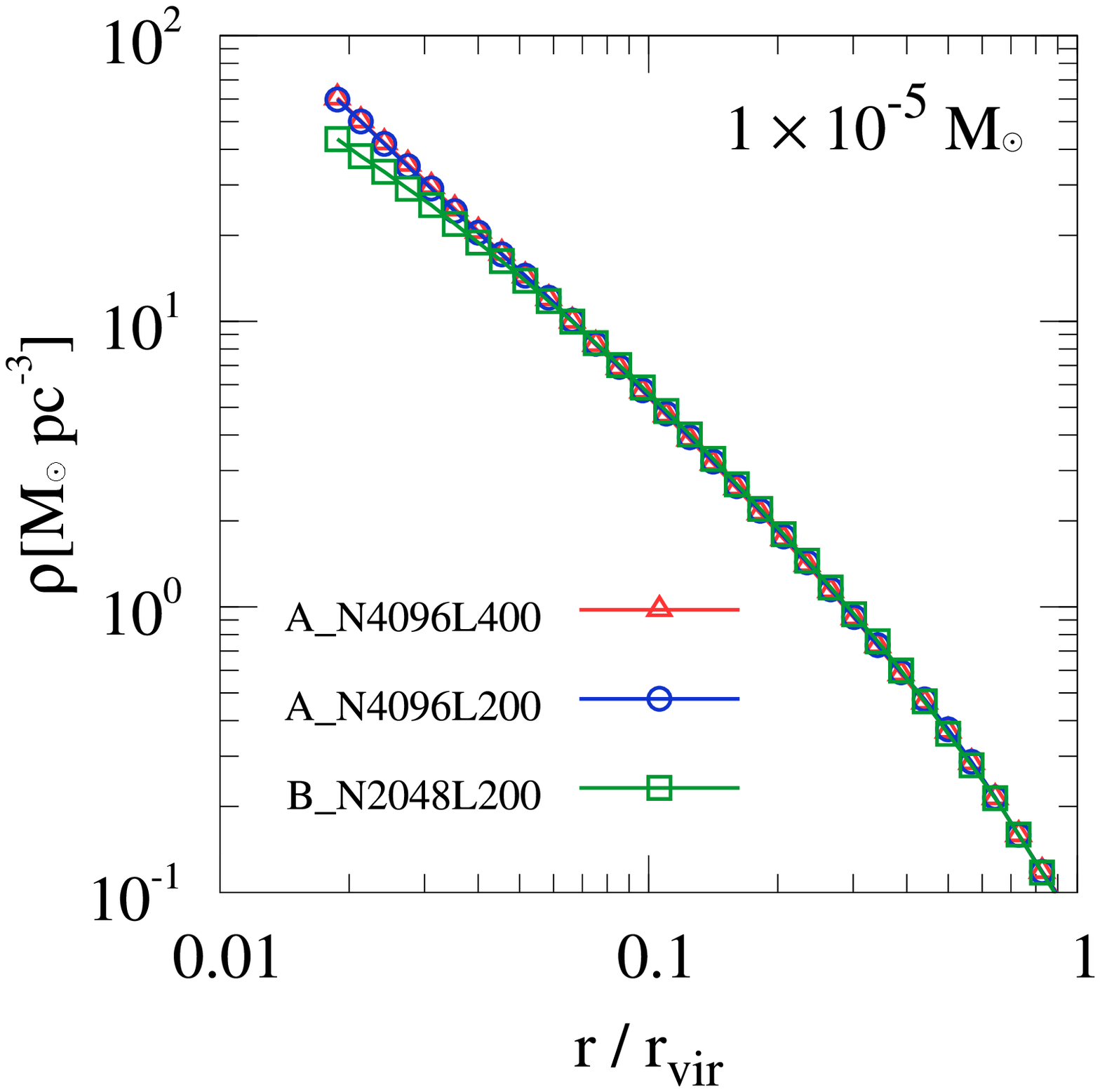} 
\includegraphics[width=5.3cm]{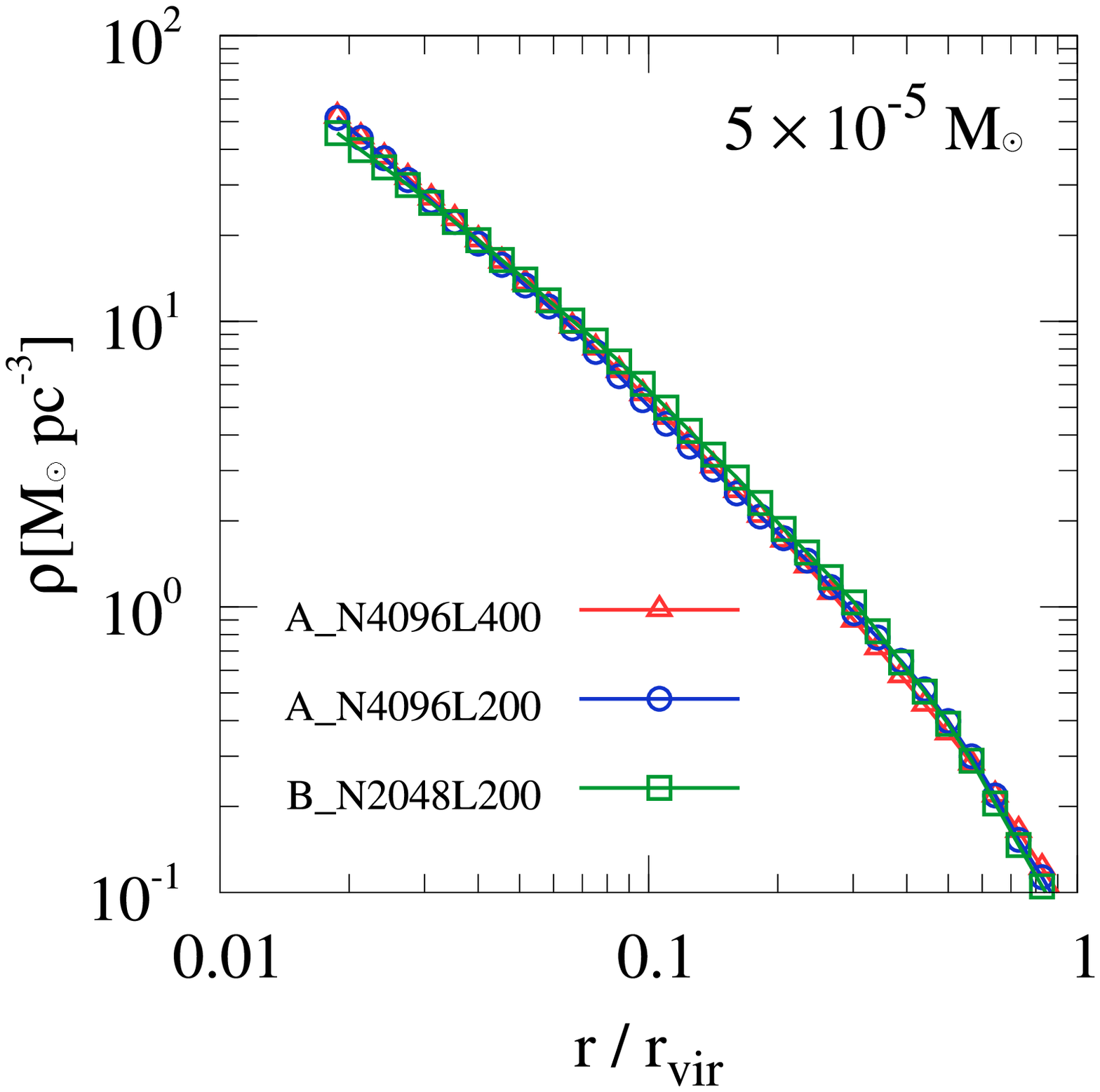} 
\includegraphics[width=5.3cm]{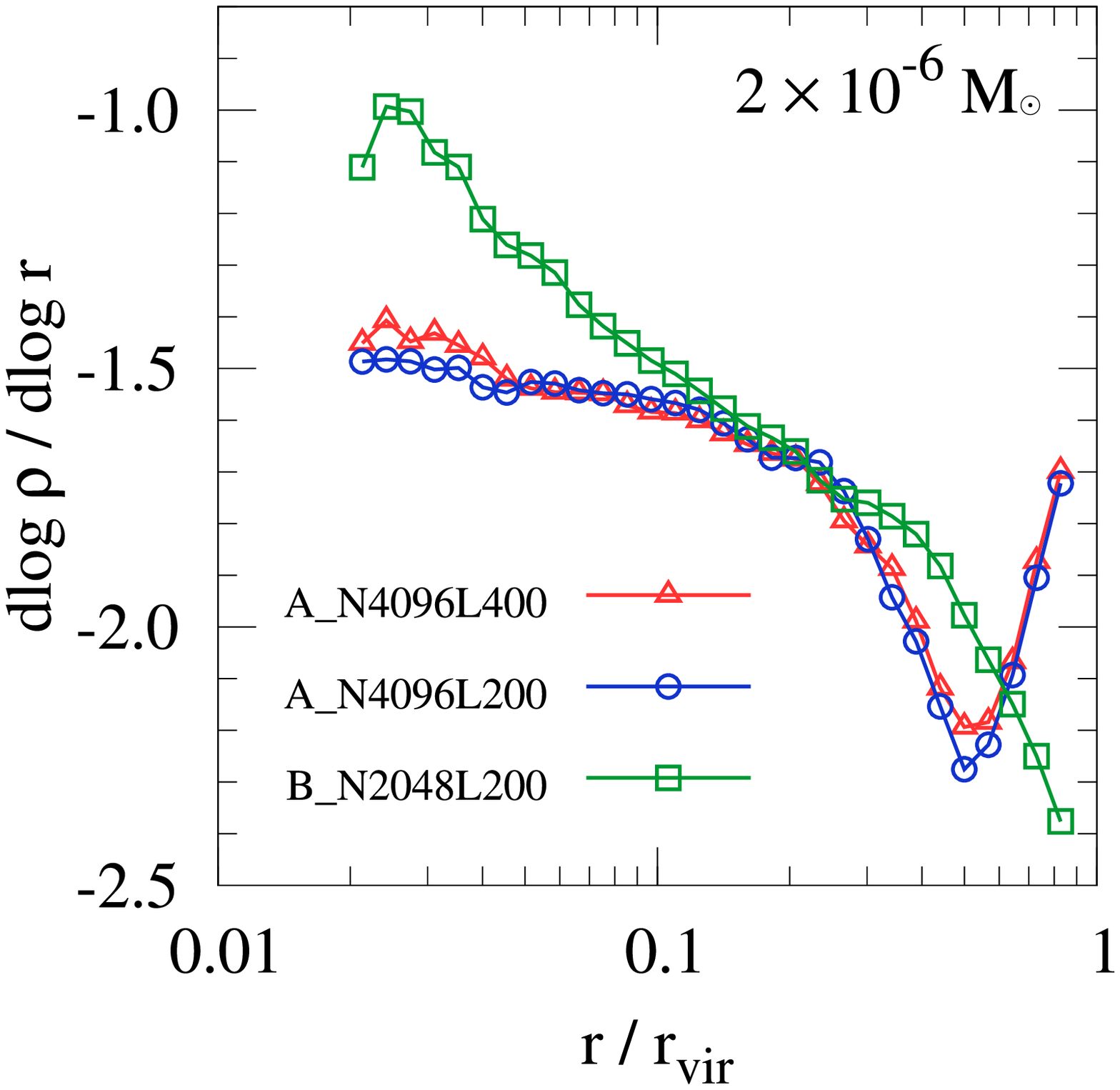} 
\includegraphics[width=5.3cm]{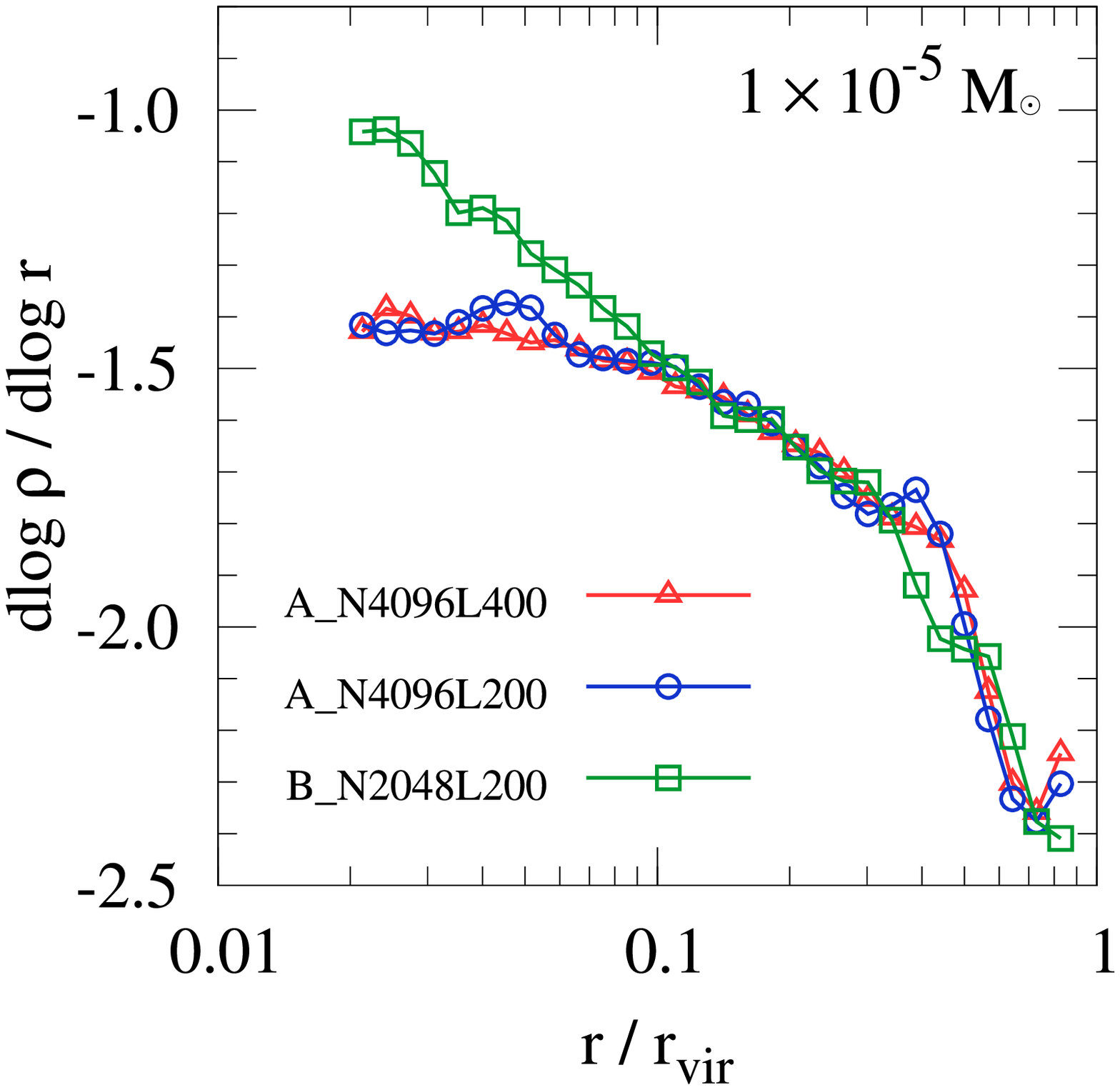} 
\includegraphics[width=5.3cm]{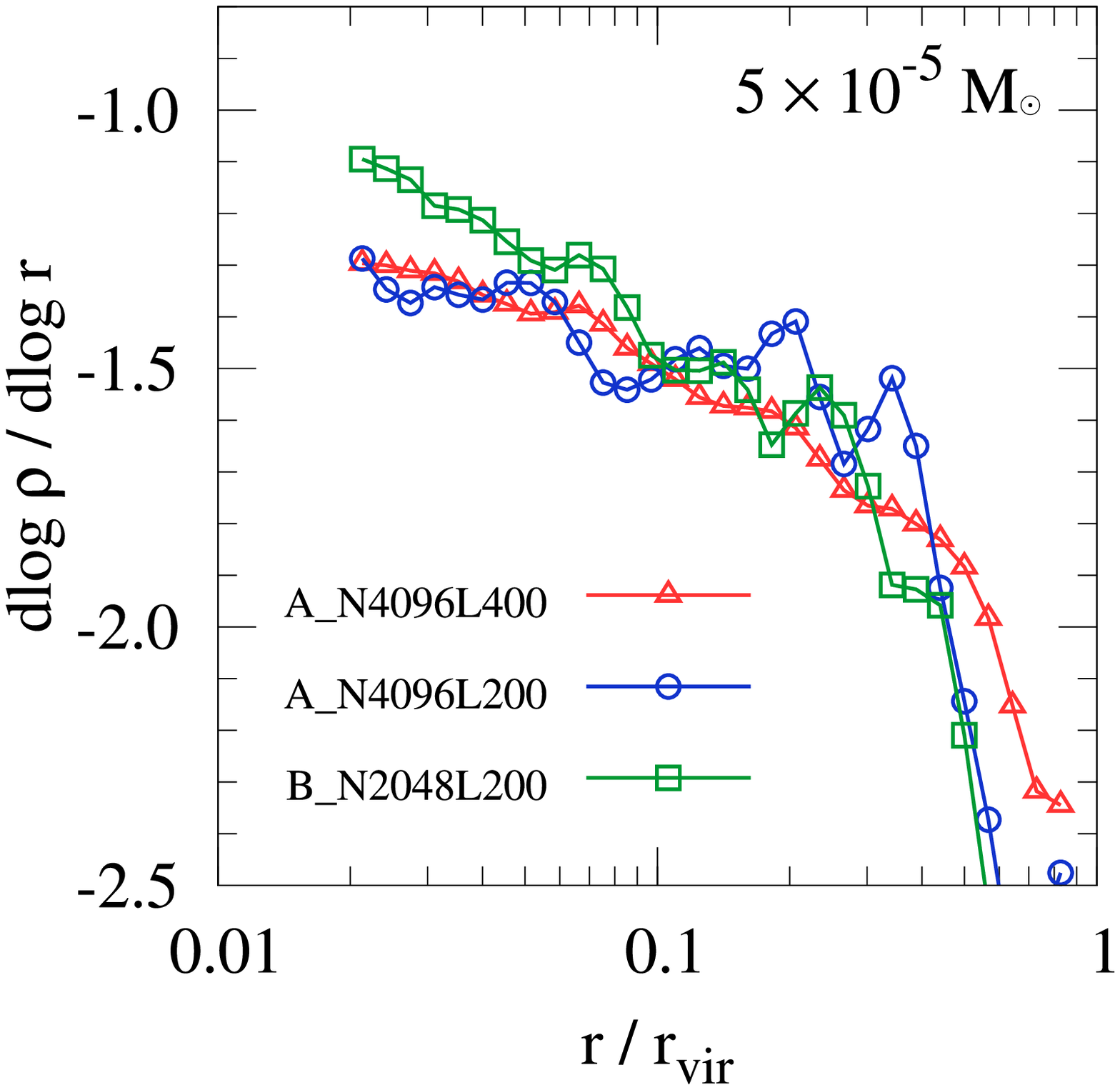} 
\caption{ 
Stacked radial density profiles (top panel) and slopes of
 profiles (bottom panel) at $z=32$ for three simulations as a
 function of radius (normalized by the virial radius).  
}
\label{fig:prof_comp}
\end{figure*}

We calculated the spherically averaged radial density profile of each
halo within the range $0.02 \le r/r_{\rm vir} \le 1.0$, divided into
32 logarithmically equal intervals.  Each density profile deviates to
varying extent from the average density profile, mainly because
subhalos exist in the halos.  To minimize this effect and obtain
proper average radial density profiles of halos with a wide mass
distribution, we stacked the profiles of similar-mass halos.

Halos resolved by a small number of particles suffer from numerical
artifacts introduced by two-body relaxation.  First, we evaluate the
minimum mass that yields a reliable density profile.  Figure
\ref{fig:prof_conv} shows the normalized stacked
density differences at $z=32$ between the \Af\ simulation and the
\As\ simulation for halos of different masses ($5 \times 10^{-7}
M_{\odot}$, $1 \times 10^{-6} M_{\odot}$ and $2 \times 10^{-6}
M_{\odot}$).  The density profile of the higher mass halo ($2 \times
10^{-6} M_{\odot}$) is almost identical in both simulations.  The
difference is below 5\% within the radial range $0.02 \le r/r_{\rm
  vir} \le 1.0$.  For the lower-mass halos ($5 \times 10^{-7}
M_{\odot}$ and $1 \times 10^{-6} M_{\odot}$), poorer agreement between
the simulations is observed across most ranges.  The \Af\ simulation
yields smaller densities than the \As\ simulation from 
$\sim0.05r/r_{\rm vir}$ and the difference is comparable to 10\% at
$0.02r/r_{\rm vir}$.

The mass resolution in the \As\ simulation is eight times higher than
in the \Af\ simulation.  We infer that these density decreases are
numerical artifacts introduced by two-body relaxation.  Hereafter, we
conservatively use the stacked density profiles of halos with the
masses larger than $2 \times 10^{-6} M_{\odot}$ for the
\Af\ simulation and the \Bf\ simulation since both mass resolution are
same.

Figure \ref{fig:prof_comp} shows the stacked radial density profiles
of three simulations at $z=32$ for three different mass bins, $2
\times 10^{-6} M_{\odot}$, $1 \times 10^{-5} M_{\odot}$, and $5 \times
10^{-5} M_{\odot}$.  The density profiles of the \Af\ and
\As\ simulations show excellent agreement.  However, large differences
appear between the simulations with the cutoff (\Af\ and \As) and that
without the cutoff (\Bf).  For halos of $2 \times 10^{-6} M_{\odot}$,
the central cusps in the density profiles are substantially steeper
when the cutoff is imposed.  These results are consistent with those
of earlier works \citep{Ishiyama2010, Anderhalden2013}.  The effect of
the cutoff is reduced in halos of higher mass ($1 \times 10^{-5}
M_{\odot}$ and $5 \times 10^{-5} M_{\odot}$).  These results are
highlighted in the bottom panel of Figure \ref{fig:prof_comp}, which
plots the local logarithmic slopes of the density profiles.

\begin{figure*}
\centering 
\includegraphics[width=5.3cm]{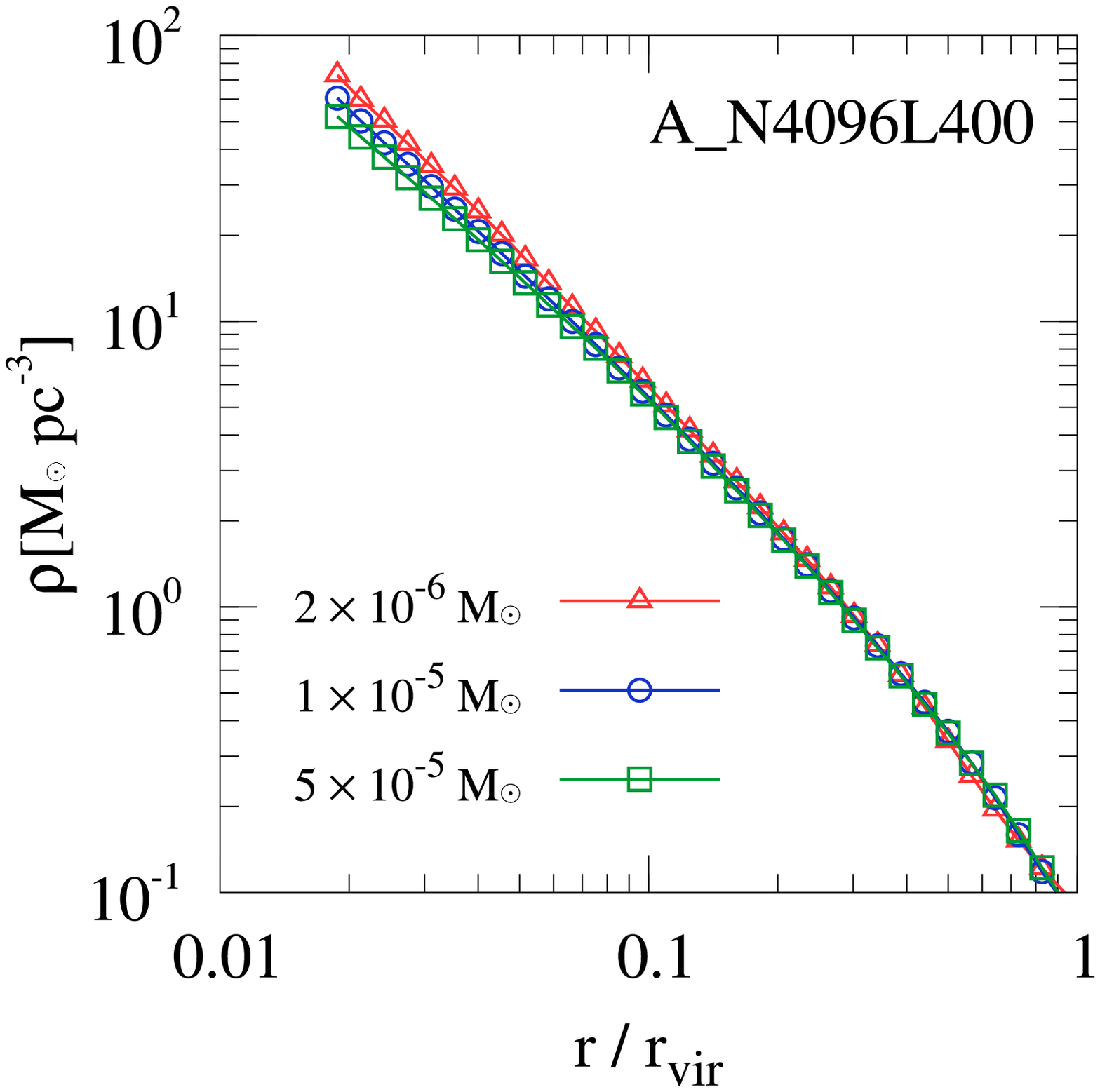} 
\includegraphics[width=5.3cm]{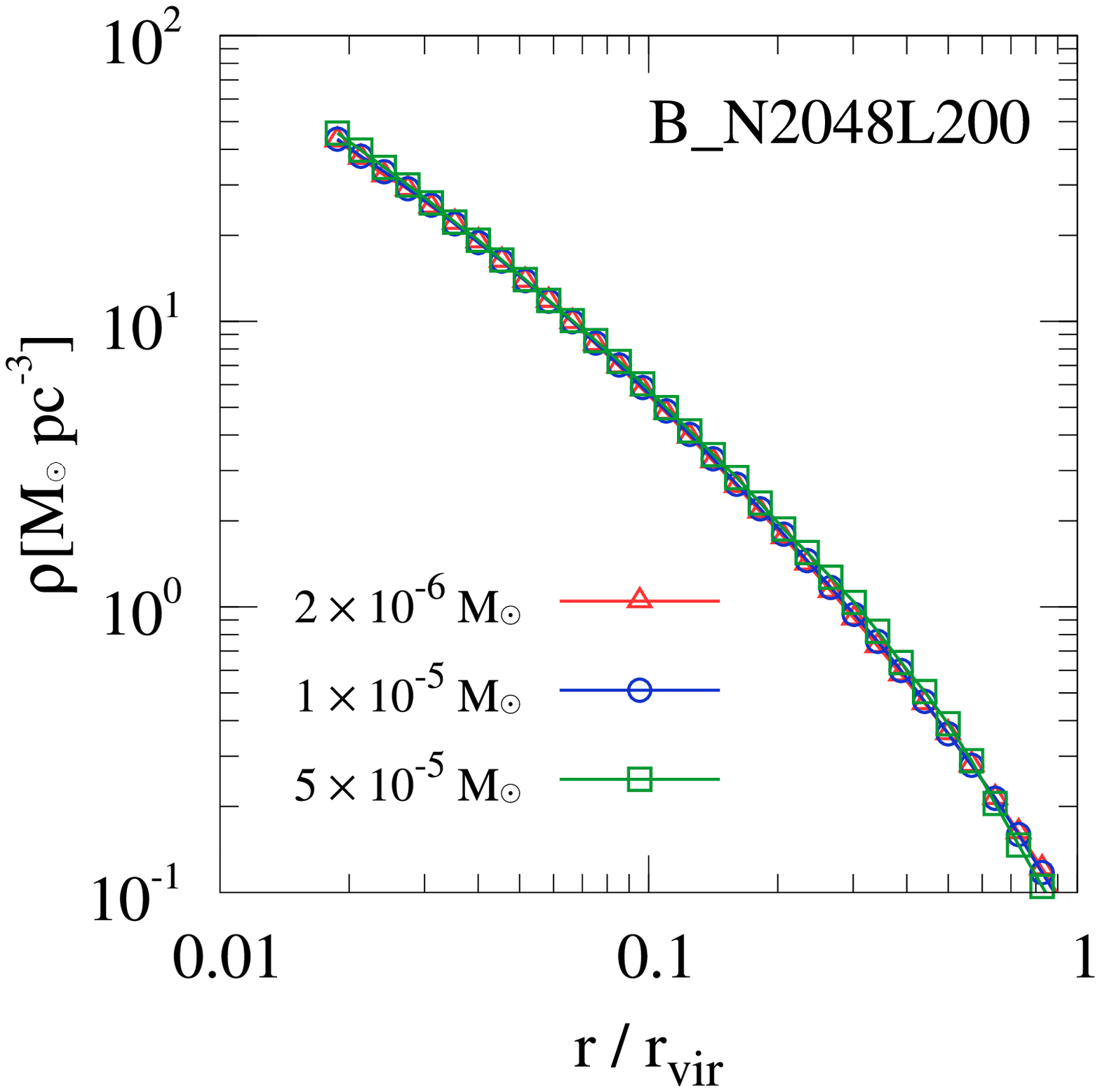} \\
\includegraphics[width=5.3cm]{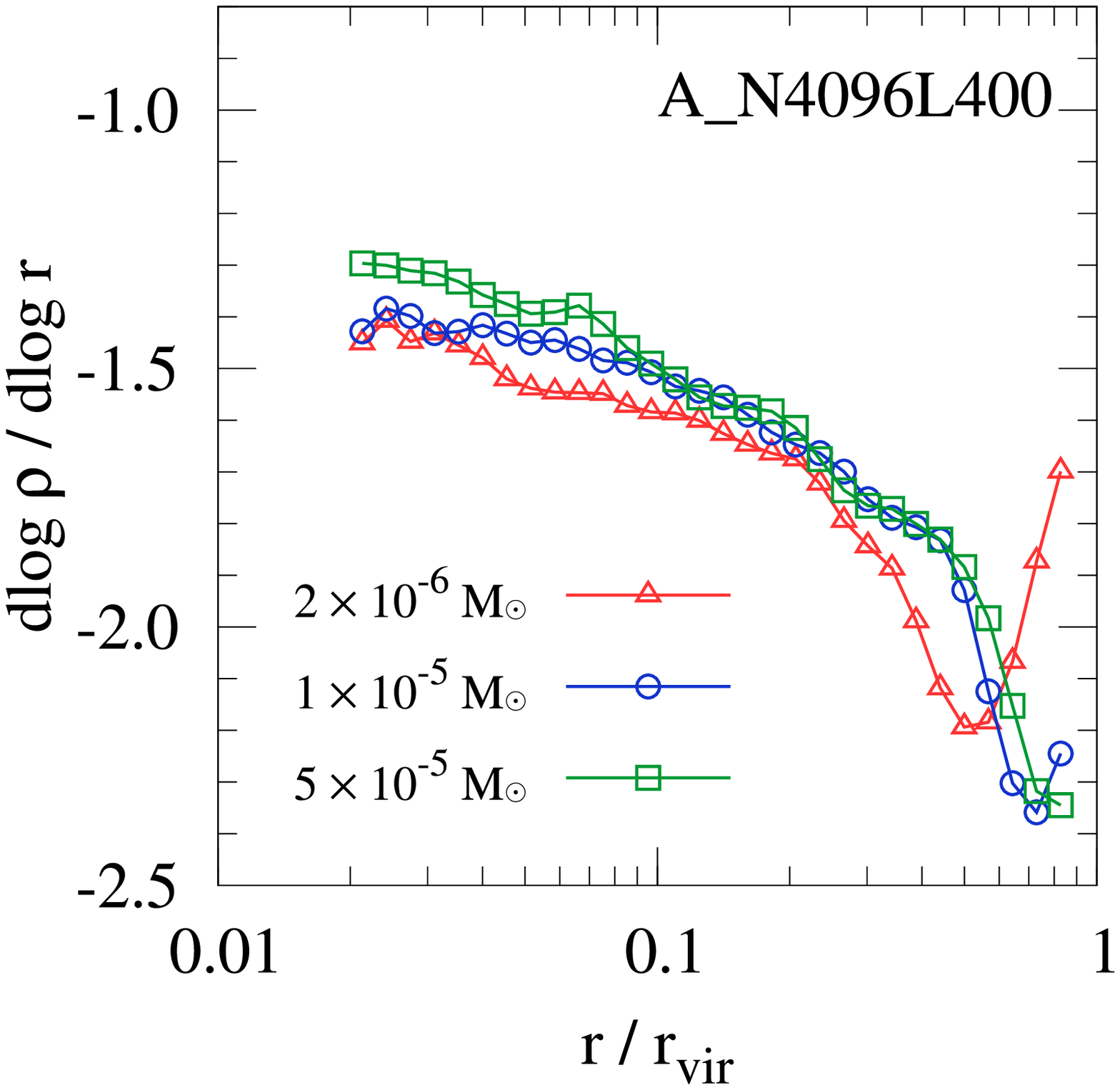} 
\includegraphics[width=5.3cm]{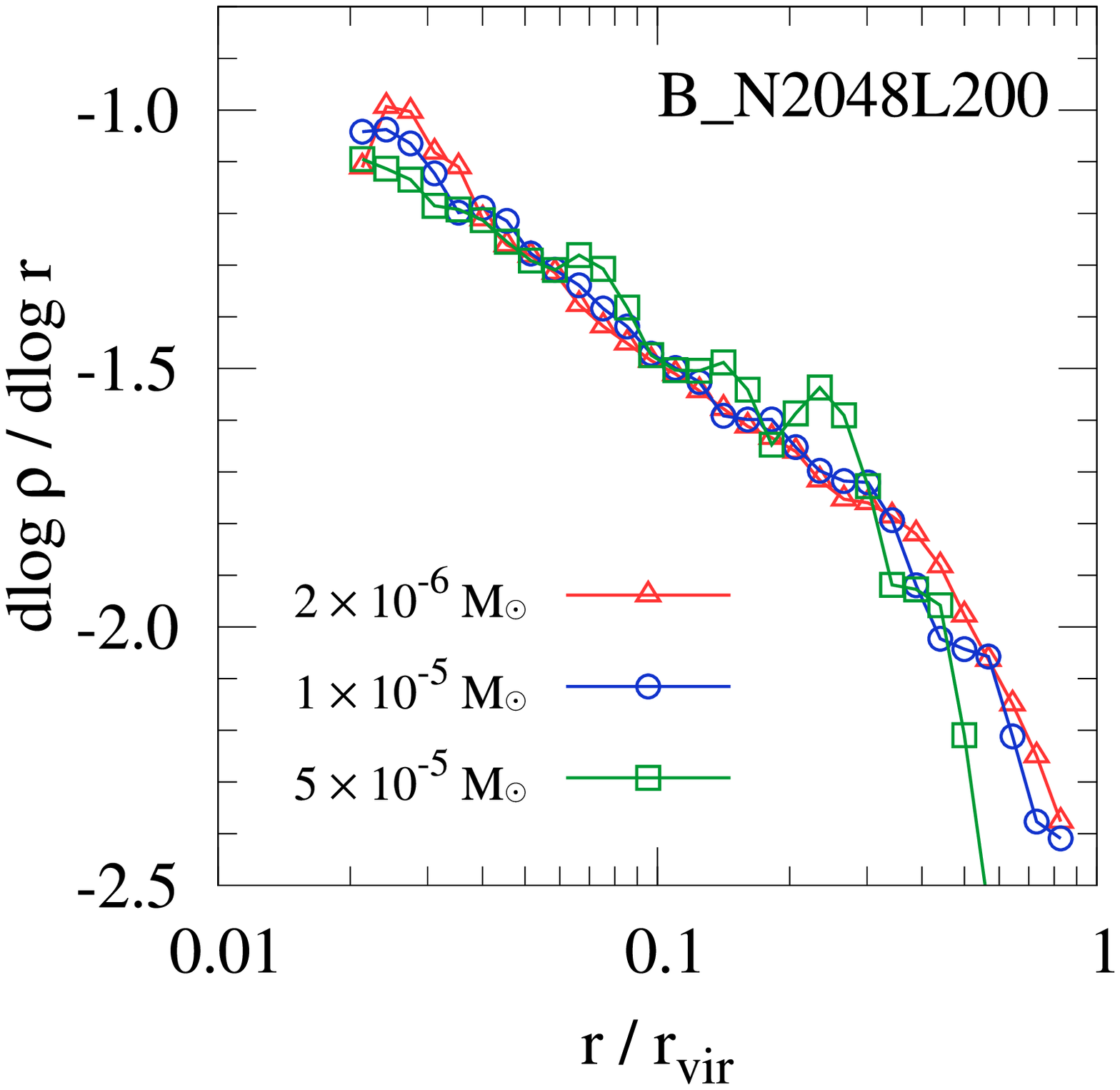} 
\caption{ 
Stacked radial density profiles (top panel) 
and slope of profile (bottom panel) for \Af\ and \Bf\ simulations
as a function of radius (normalized by the virial radius).
}
\label{fig:prof}
\end{figure*}

These surprising results are further highlighted in Figure
\ref{fig:prof}.  The stacked density profiles are plotted as in Figure
\ref{fig:prof_comp}, but each panel shows the density profiles of
three different mass bins taken from the \Af\ and
\Bf\ simulations. With the cutoff imposed, the slope of the central
cusp gradually shallows at higher masses.  For halos of $2 \times
10^{-6} M_{\odot}$, the central slope is around $-1.5\sim-1.4$,
consistent with earlier reports \citep{Ishiyama2010,
  Anderhalden2013}.  For $5 \times 10^{-5} M_{\odot}$ halos, the
central slope is reduced to around $-1.3$.  Remarkably, the
central slopes of density profiles are almost flat and tend toward a
constant value at $r/r_{\rm vir} \le \sim0.04$.

Without the cutoff, the results are dramatically different. As seen in
the right panels of Figure \ref{fig:prof}, the slopes of the density
profile become shallower at inner radii and do not converge to a
single power law.  
The shape of the inner density profile appears to be independent of halo mass,
consistent with the results of cosmological simulations for
galaxy-sized and cluster-sized halos \citep[e.g.,][]{Fukushige2004,
  Stadel2009, Ishiyama2013}.

Why the inner density profile becomes shallower towards the center in 
simulations without cutoff
and of larger halos is not currently 
understood, and is not pursued further in this
paper.  In \S \ref{sec:profile_evo}, we consider the physical origin
of steep density cusps observed in the simulations with the cutoff.

%%%%%%%%%%%%%%%%%%%%%%%%%%%%%%%%%%%%%%%%%%%%%%%%%%%%%%%%%%%%%%%%%%%%%%%%%%%%%%
%%%%%%%%%%%%%%%%%%%%%%%%%%%%%%%%%%%%%%%%%%%%%%%%%%%%%%%%%%%%%%%%%%%%%%%%%%%%%%

\subsection{Fitting Microhalo Density Profiles}\label{sec:profile2}

\subsubsection{Shape of the Inner Density Profiles}

\begin{figure}
\centering 
\includegraphics[width=4.2cm]{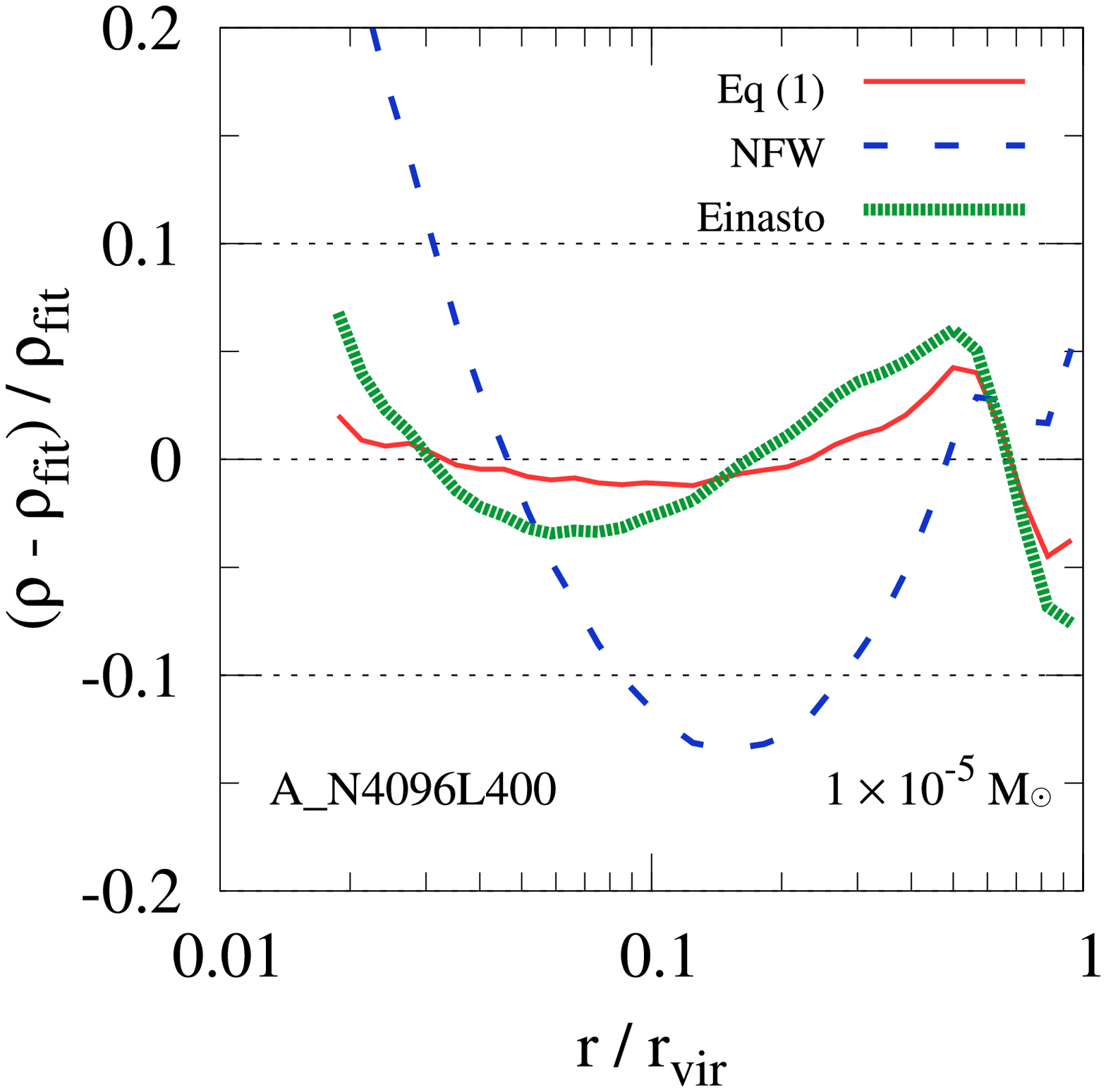} 
\includegraphics[width=4.2cm]{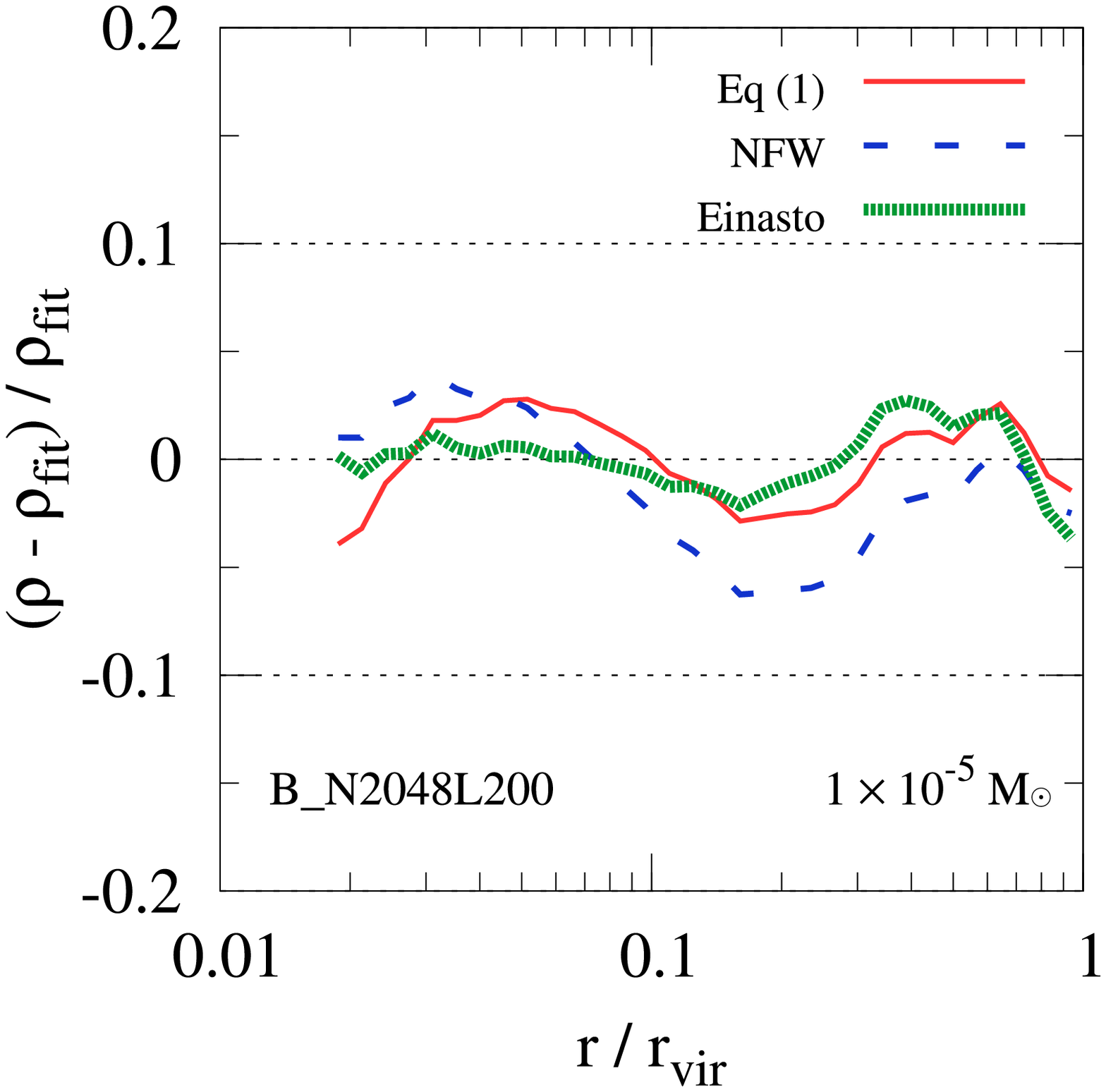} 
\caption{ 
Residuals of stacked radial density profiles obtained
from the simulations, relative to the three fitting functions.
}
\label{fig:prof_fit}
\end{figure}

\begin{figure}
\centering 
\includegraphics[width=8.5cm]{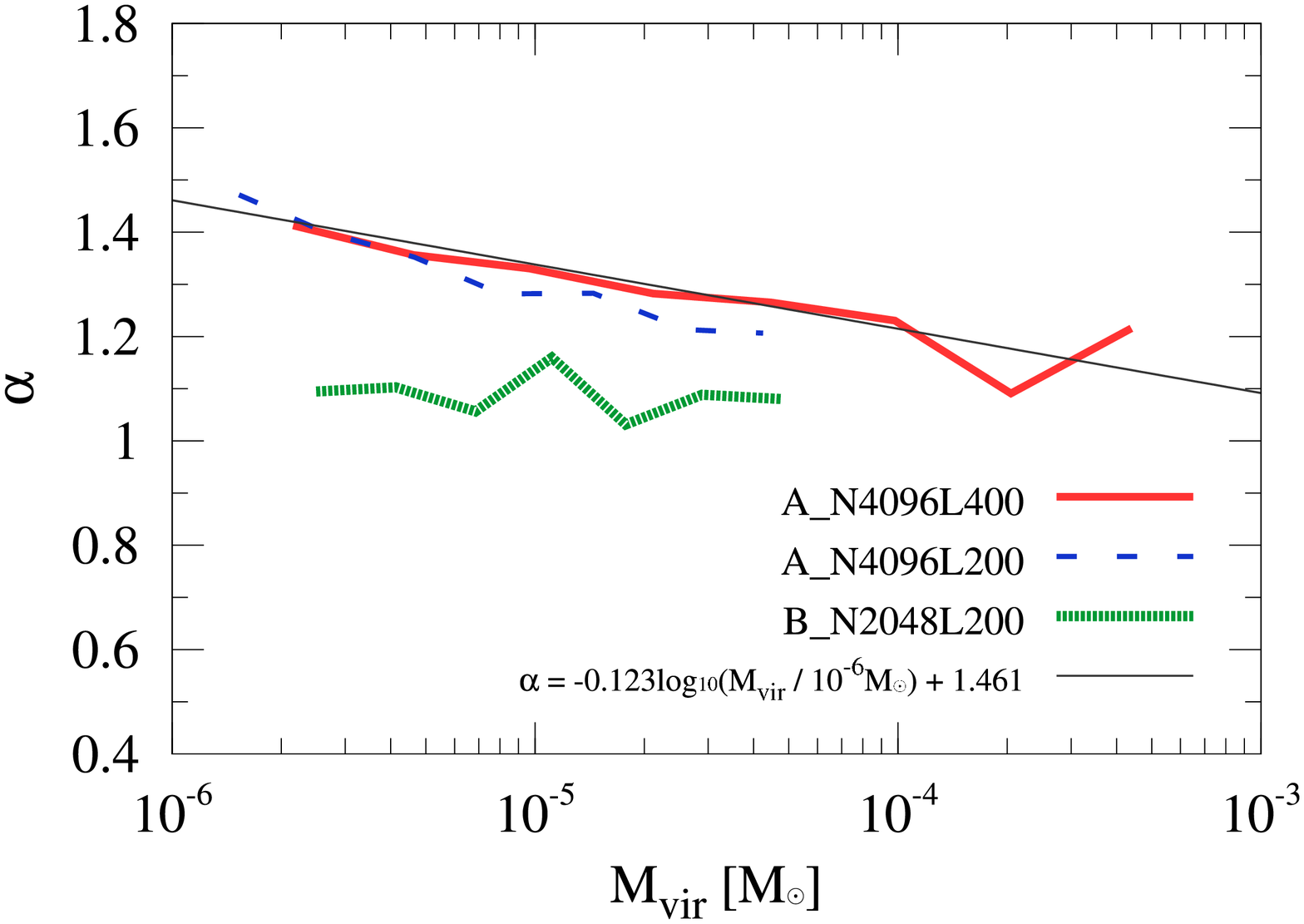} 
\caption{Slopes of stacked density profiles $\alpha$ as a function of
  the halo virial mass $M_{\rm vir}$. These slopes are derived by
  fitting the profiles to a double power law function (Equation
    (\ref{eq:doublepower}) in the text).  The fittings are performed
  with the data divided into 32 logarithmically equal bins from
  $0.02r_{\rm vir}$ to $r_{\rm vir}$.  Black solid line shows the best
  fit power law function (Equation (\ref{eq:fita}) in the text).}
\label{fig:m-alpha}
\end{figure}

To quantify the simulated density structures, we fitted the density
profiles to fitting functions.  Two popular functions used in
high-resolution cosmological simulations are the NFW and the Einasto
profiles.  In these fitting functions, the central slope is
approximately $-1$ or less.  Therefore, whether these functions can
adequately describe the density profiles obtained in the simulations
with the cutoff is uncertain.

Besides the NFW and Einasto profiles, we 
tried a double power 
law function, given by 
\begin{eqnarray}
\rho(r) = \frac{\rm \rho_0}{ (r/r_{\rm s})^{\alpha} (1+r/r_{\rm s})^{(3-\alpha)}}. 
\label{eq:doublepower}
\end{eqnarray} 
This function, which has been fitted to density profiles in previous works
\citep[e.g.,][]{Diemand2004b, Anderhalden2013}, 
is identical to the NFW profile when $\alpha=1$. 
Because the inner slope can vary, we expect that the density 
profiles obtained in the simulations with the cutoff can be precisely described 
by this function.

As shown in Figure \ref{fig:prof_fit}, all three of these fitting
functions are well fitted to the density profiles simulated without
the cutoff, consistent with previous works.  The NFW profile yields a
marginally worse fit than the other two functions, possibly because
the other functions have an additional shape parameter $\alpha$.  On
the other hand, the profile of the cutoff model is well-fitted only to
Equation (\ref{eq:doublepower}).  As expected, the NFW profile
improperly describes the density profile since it forces the central
slope to be $-1$.  The error in the Einasto profile is acceptable, but
function (\ref{eq:doublepower}) provides a clearly superior fit.
Hereafter, we use this double power function to fit the density
profile obtained in simulations.

Figure \ref{fig:m-alpha} shows the slopes of the stacked density profiles
$\alpha$ at $z=32$ as functions of the halo virial mass.  
Only mass ranges containing more than 15 halos are plotted. 
As expected, the
slopes are almost constant for the \Bf\ simulation.
The value of $\alpha$ ($\sim1.1$) agrees well with 
that of the NFW profile. 
Simulations \Af\ and \As\ yield very similar results. 
In these simulations, the slope $\alpha$ is considerably
larger than in \Bf\, as seen in Figures \ref{fig:prof_comp} and \ref{fig:prof}.
The slope difference reduces as the halo mass increases. 
Note that at the cutoff scale ($\sim10^{-6}M_{\odot}$), 
the slope $\alpha$ is around $1.4\mbox{\scriptsize --}1.5$, 
which is consistent with previous works such as 
\citet[][$\alpha=1.5$]{Ishiyama2010} and 
\citet[][$\alpha=1.3\mbox{\scriptsize --}1.4$]{Anderhalden2013}.
The power law functions that best fits the relation between mass and $\alpha$ is
\begin{eqnarray}
\alpha = -0.123 \log( M_{\rm vir} / 10^{-6}M_{\odot}) + 1.461. \label{eq:fita}
\end{eqnarray}
This function (solid black line in Figure \ref{fig:m-alpha})
matches the numerical results quite accurately. 

\begin{figure}
\centering 
\includegraphics[width=8.5cm]{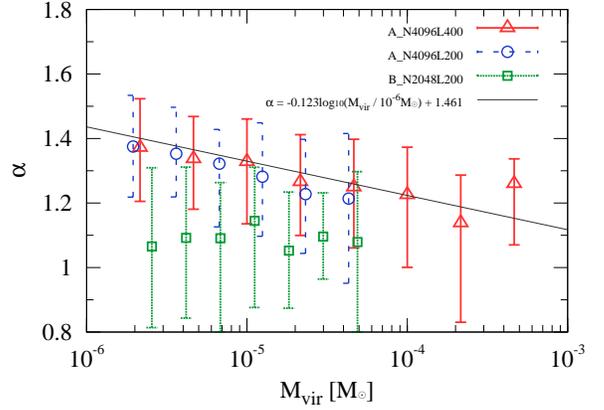}
\caption{Slope of the density profile of each halo $\alpha$ plotted
  against the halo virial mass $M_{\rm vir}$. Circles, triangles and
  squares show the median value in each mass bin. Whiskers are the
  first and third quantiles.
Black solid line is the best fit power law function
(Equation (\ref{eq:fita}) in the text).}
\label{fig:m-alpha_bin}
\end{figure}

\begin{figure}
\centering 
\includegraphics[width=8.5cm]{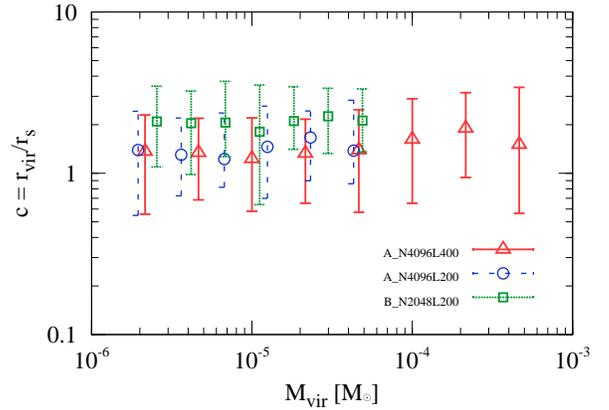} 
\caption{Concentration of the density profile of each halo $r_{\rm
    vir}/r_{\rm s}$   plotted against the halo virial mass $M_{\rm
    vir}$ at $z=32$. 
  Circles, triangles, and squares show the median value in
  each mass bin. Whiskers are the first and third quantiles.  }
\label{fig:m-c_bin}
\end{figure}

To visualize the scatter in the density profiles,
we fitted the profile of each halo to Equation (\ref{eq:doublepower}) 
and calculated the median and scatter in each mass bin. 
The median, and the first and third quantiles of the shape parameter
$\alpha$ are plotted against halo virial mass in Figure
\ref{fig:m-alpha_bin}.  The two simulations with different resolutions
give similar results.  The median accurately matches the fitting
function derived from the stacked density profile.  Regardless of halo
mass, the first and third quantiles deviate by less than 20\% in the
\Af\ simulation.  Clearly, the \Bf\ simulation generates more scatter
than the \Af\ simulation.

\subsubsection{Microhalo Concentrations}\label{sec:concentration}

Figure \ref{fig:m-c_bin} shows the median, first and third quantiles
of the concentration parameter $c=r_{\rm vir}/r_{\rm s}$ as a function
of halo virial mass.  Note that the concentration parameter is defined
differently than the NFW profile, since another fitting function is
used.  Remarkably, the concentration parameter in both models is
nearly independent of halo mass over the range shown in Figure
\ref{fig:m-c_bin}.  The median concentration in the cutoff model is
$1.2\mbox{\scriptsize --}1.7$, increasing to $1.8\mbox{\scriptsize
  --}2.3$ without the cutoff.

These results differ from what we see in larger halos
(dwarf-galaxy-sized to cluster-sized halos) at lower redshifts
(typically less than $z=5$).  The concentrations of these halos weakly
depend on the halo mass, and have been fitted to many simple single
power law functions \citep[e.g.,][]{Neto2007}.  
Since the slope of the power spectrum of initial density fluctuations 
approaches $-3$ for the small mass limit,
the relation weakens as the halo mass decreases.
There are models that connect 
concentrations to the rms of the linear density fluctuation field 
and reproduce the mass--concentration relation 
derived from cosmological simulations
\citep[e.g.,][]{Bullock2001,Maccio2008, Prada2012, Sanchez-Conde2013}.

Whether these fitting functions 
can be extrapolated to halos near the 
free streaming scale is worthy of investigation.  Here, we present the 
first direct comparison using simulation results. In most previous studies, 
the fitting function is derived by applying the NFW profile to 
the simulated density profile.  Our simulations robustly show that 
density profiles in the cutoff simulations do not follow the NFW profile 
and are best quantified by Equation (\ref{eq:doublepower}). Increasing 
the shape parameter $\alpha$ would decrease the concentration by 
increasing the scale radius $r_{\rm s}$.  The dependence of $\alpha$ on 
the halo mass could also induce complex behavior of $r_{\rm s}$ shifts. 
Therefore, our results are not directly comparable with those reported 
elsewhere. 

To enable an indirect comparison, we converted the concentration $c$
of Equation (\ref{eq:doublepower}) to that of the NFW profile $c_{\rm
  NFW}$ at $z=0$ by a method adopted in previous works
\citep[e.g.,][]{Ricotti2003, Anderhalden2013}.  For example, when
$\alpha=1.5,1.4,1.3$ in Equation (\ref{eq:doublepower}), the
equivalent NFW concentration $c_{\rm NFW}=2.0c, 1.67c, 1.43c$.  Using
this relation and the mass--shape relation of Equation (\ref{eq:fita}),
we converted the concentration in each halo mass of the cutoff
simulations to its equivalent in the NFW profile.  In the no cutoff
simulation, where $\alpha$ was constant and equal to $1.1$, $c_{\rm
  NFW}=1.11c$.  Finally, to obtain the concentration at $z=0$, we
assumed that both the virial radius and the concentration are scalable
to $z=0$ by multiplying $(1+z)$ \citep{Bullock2001}.

\begin{figure}
\centering 
\includegraphics[width=8.5cm]{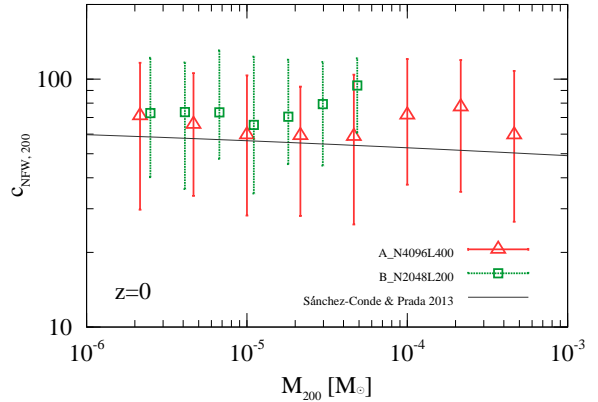} 
\caption{Converted concentration of the density profile of each halo
  $c_{\rm NFW, 200}$ plotted against the halo mass $M_{\rm 200}$.  The
  concentrations of Equation (\ref{eq:doublepower}) are converted to
  those of the NFW profile at $z=0$ as described in \S
  \ref{sec:profile2}.  Triangles, and squares show the median value in
  each mass bin. Whiskers are the first and third quantiles.  Black
  solid line is the fitting function proposed by
  \citet{Sanchez-Conde2013}. }
\label{fig:m-c_fit}
\end{figure}

Figure \ref{fig:m-c_fit} plots the converted concentration $c_{\rm
  NFW,200}$ obtained from the \Af\ and \Bf\ simulations as a function
of halo mass.  In this comparison, $M_{\rm vir}$ and $R_{\rm vir}$
are replaced by the conventionally used parameters $M_{\rm 200}$ and
$R_{\rm 200}$, respectively, in which the spherical overdensity is 200
times the critical density.  The median concentration in the cutoff
model ranges $60\mbox{\scriptsize --}70$.  The black solid line is the
function proposed by \citet{Sanchez-Conde2013} based on the results of
\citet{Prada2012}.  Although this fitting function can reproduce the
relation between concentration and mass, it consistently
underestimates the concentrations derived from the simulations.
However, the fitting lies between the first quantile and the median
over more than 10 orders of extrapolation.  Thus, the fitting proposed
by \citet{Sanchez-Conde2013} does not disagree with our results either
qualitatively or quantitatively.  
The toy models of \citet{Bullock2001} and \citet{Maccio2008} 
give similar values. 
However, when simple single power law functions are extrapolated 
to microhalo scales \citep[e.g.,][]{Springel2008b}, 
the resulting  concentrations are $\sim1000$,
approximately one order of magnitude higher than those yielded by
numerical simulations and the resulting annihilation 
boost factors are too high by large factors 
(as shown in \citealt{Sanchez-Conde2013}).
Our statistical studies of halos near the
cutoff scale robustly show that a single power law function cannot be
reliably extrapolated to this scale.

Previously, we simulated microhalos with masses of order $10^{-6}
M_{\odot}$ \citep{Ishiyama2010}, obtained a concentration $c_{\rm NFW,
  200}=60\mbox{\scriptsize --}70$.  Very similar median concentrations
were obtained in the present study.  Other simulation of microhalo
with masses of order $10^{-7} M_{\odot}$, \citep{Anderhalden2013}
yielded $c_{\rm NFW, 200}=94\mbox{\scriptsize --}124$.  Although these
concentrations exceed the median value of those obtained in our
simulations, they reside between the first and third quantiles.  We
emphasize that the results of previous microhalo simulations, in which
only a few halos were simulated \citep{Ishiyama2010, Anderhalden2013},
are not inconsistent with the present study.

Our results rule out single power law mass--concentration relations 
\citep[e.g.,][]{Neto2007}.
However, our simulated
concentration are slightly but systematically shifted upward 
from the fitting of \citet{Sanchez-Conde2013}.  
Whether this anomaly is caused by some limitation of the simulations (e.g.,
absence of the long-wave perturbations) or the accuracy of fitting, 
could be explored in larger box simulations.

%%%%%%%%%%%%%%%%%%%%%%%%%%%%%%%%%%%%%%%%%%%%%%%%%%%%%%%%%%%%%%%%%%%%%%%%%%%%%%
%%%%%%%%%%%%%%%%%%%%%%%%%%%%%%%%%%%%%%%%%%%%%%%%%%%%%%%%%%%%%%%%%%%%%%%%%%%%%%
\subsection{Dependence of Density Profile on Halo Formation Epoch}\label{sec:profile3}

\begin{figure*}
\centering 
\includegraphics[width=5.3cm]{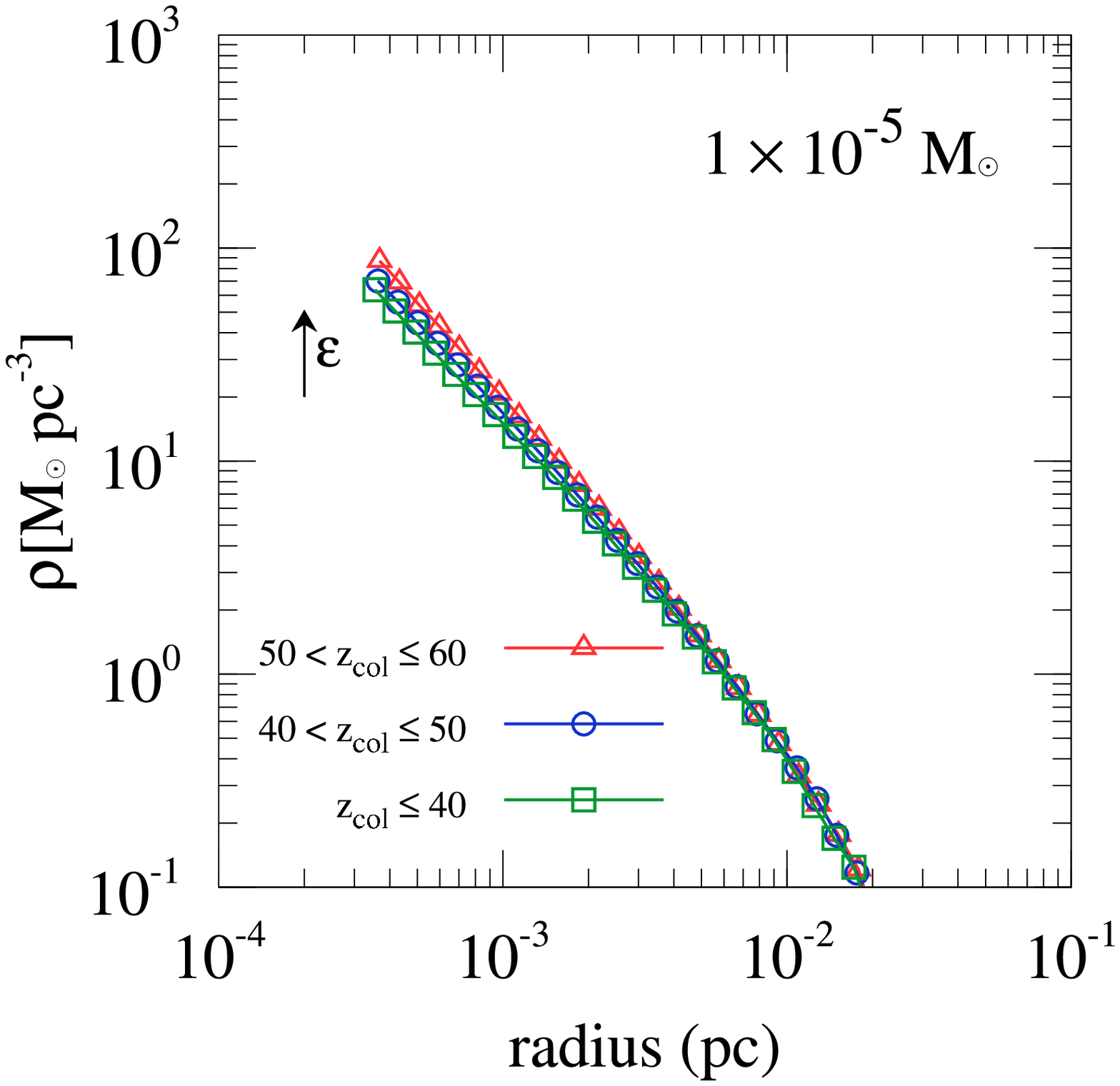}
\includegraphics[width=5.3cm]{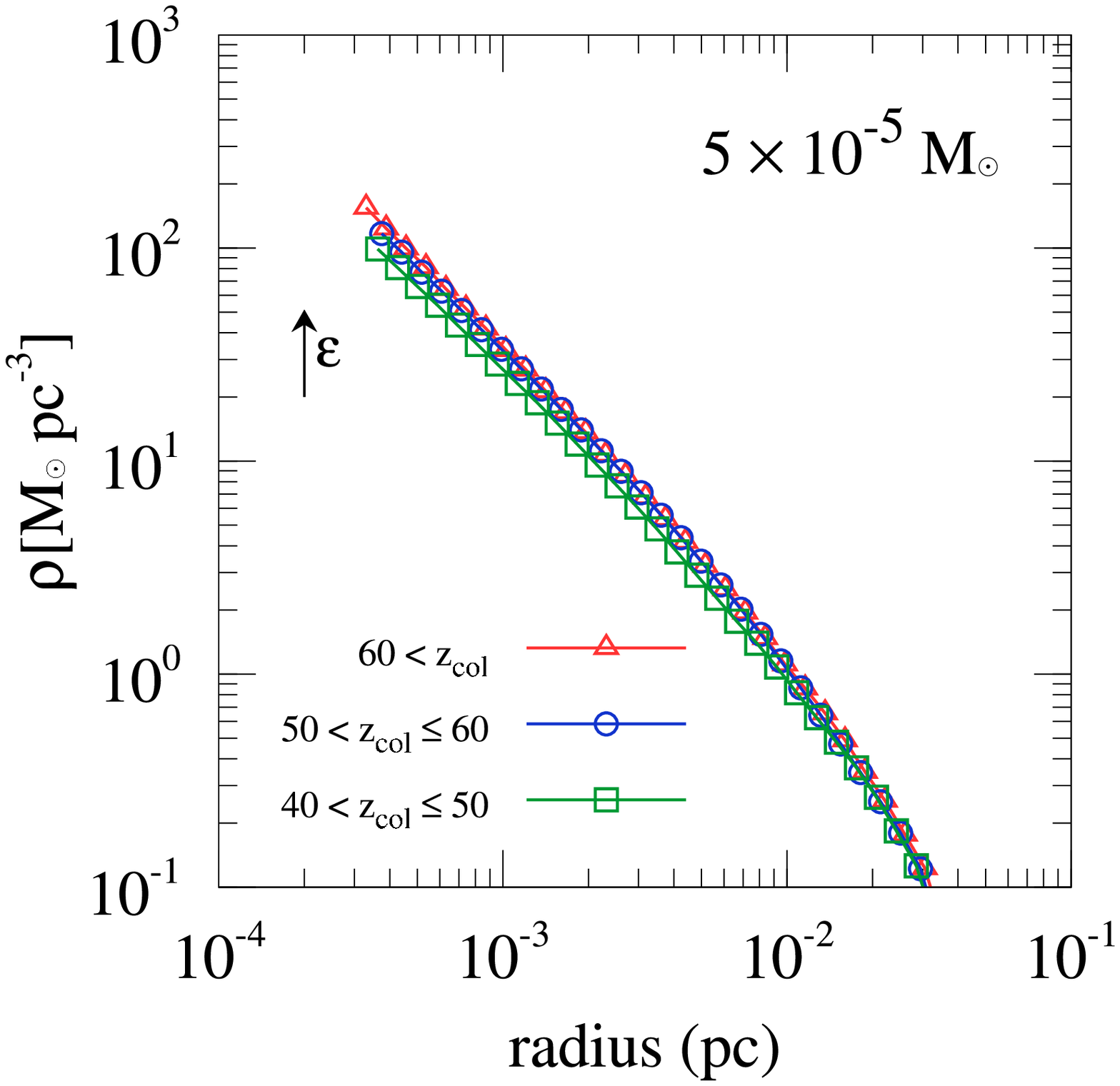}
\includegraphics[width=5.3cm]{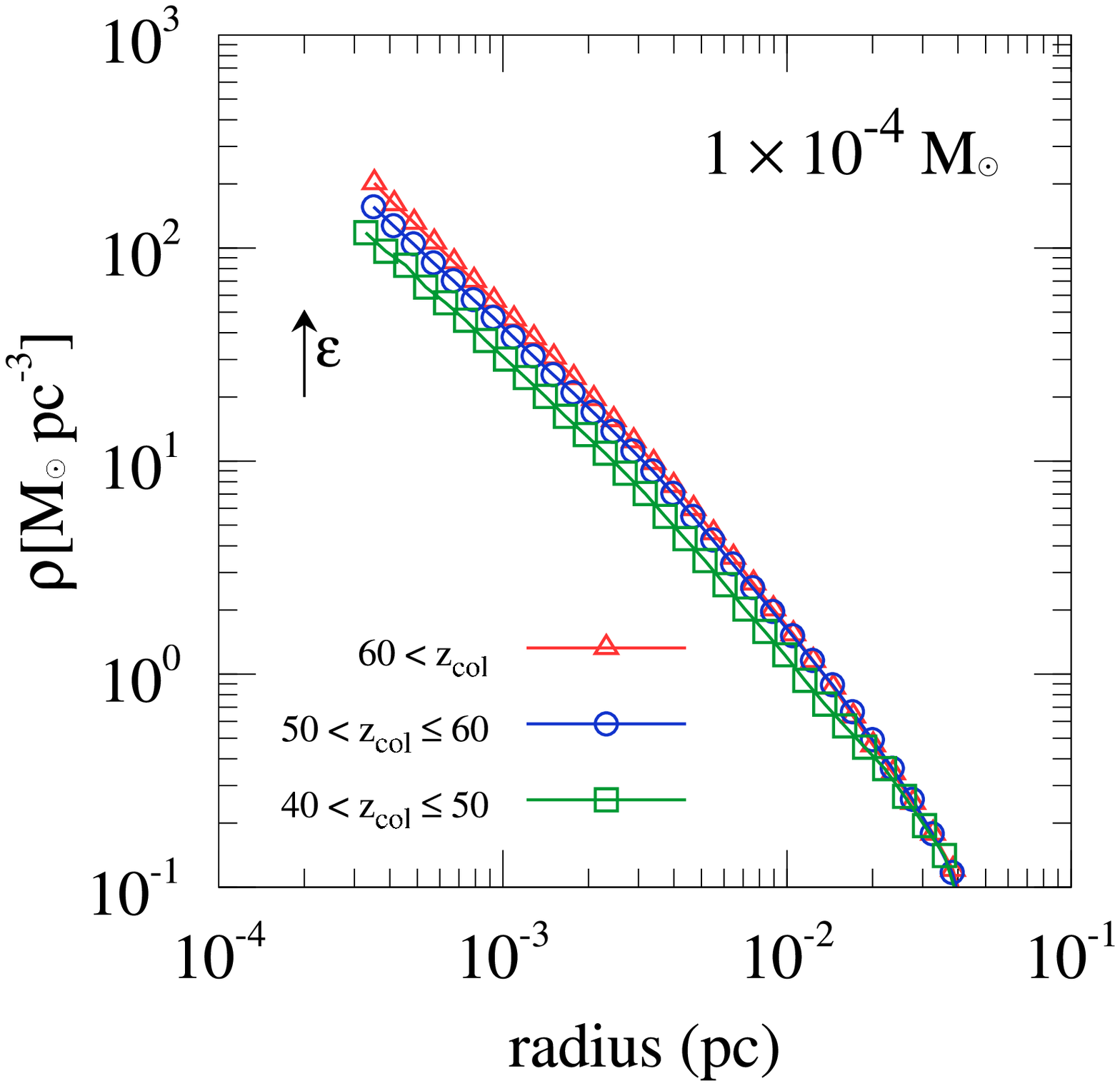}
\includegraphics[width=5.3cm]{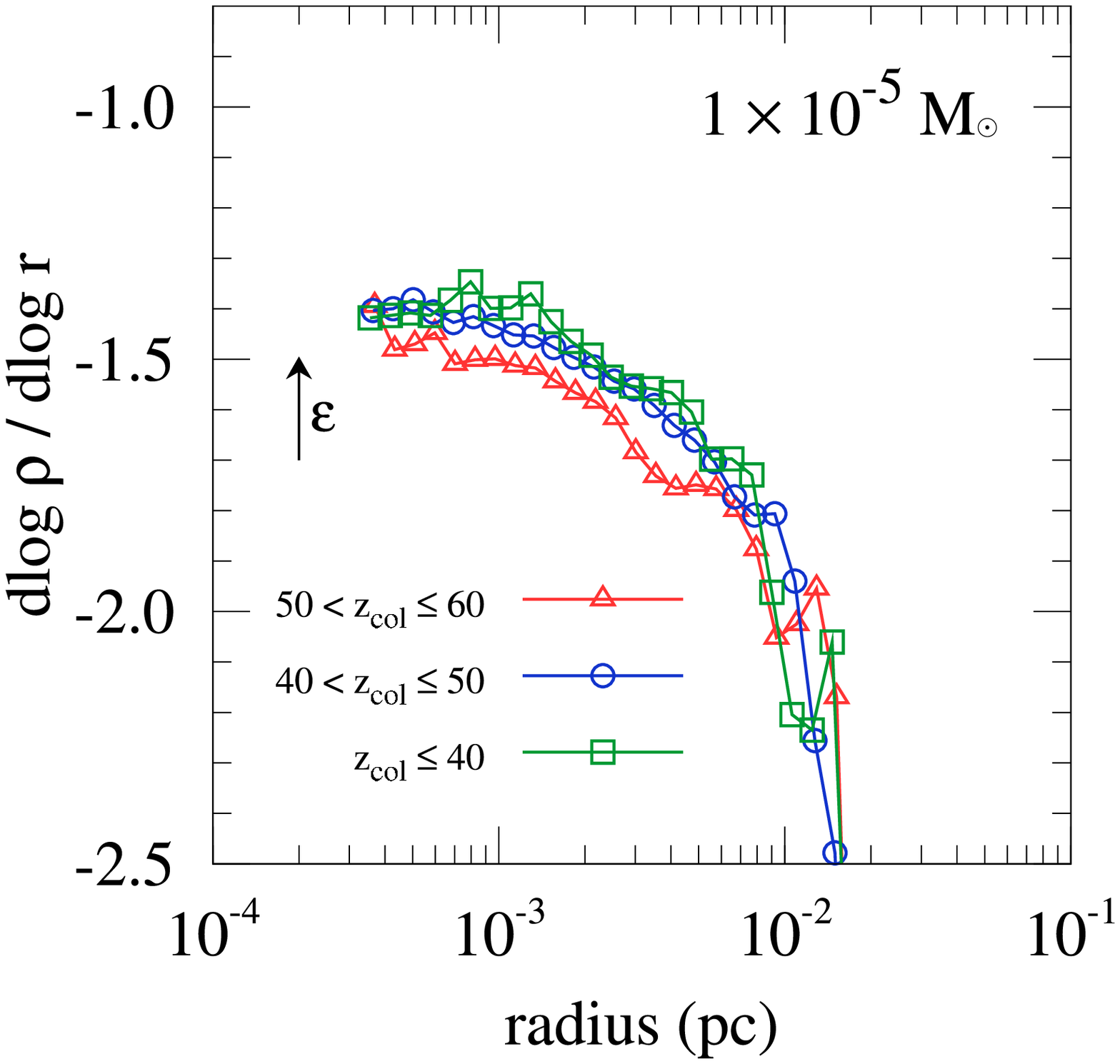}
\includegraphics[width=5.3cm]{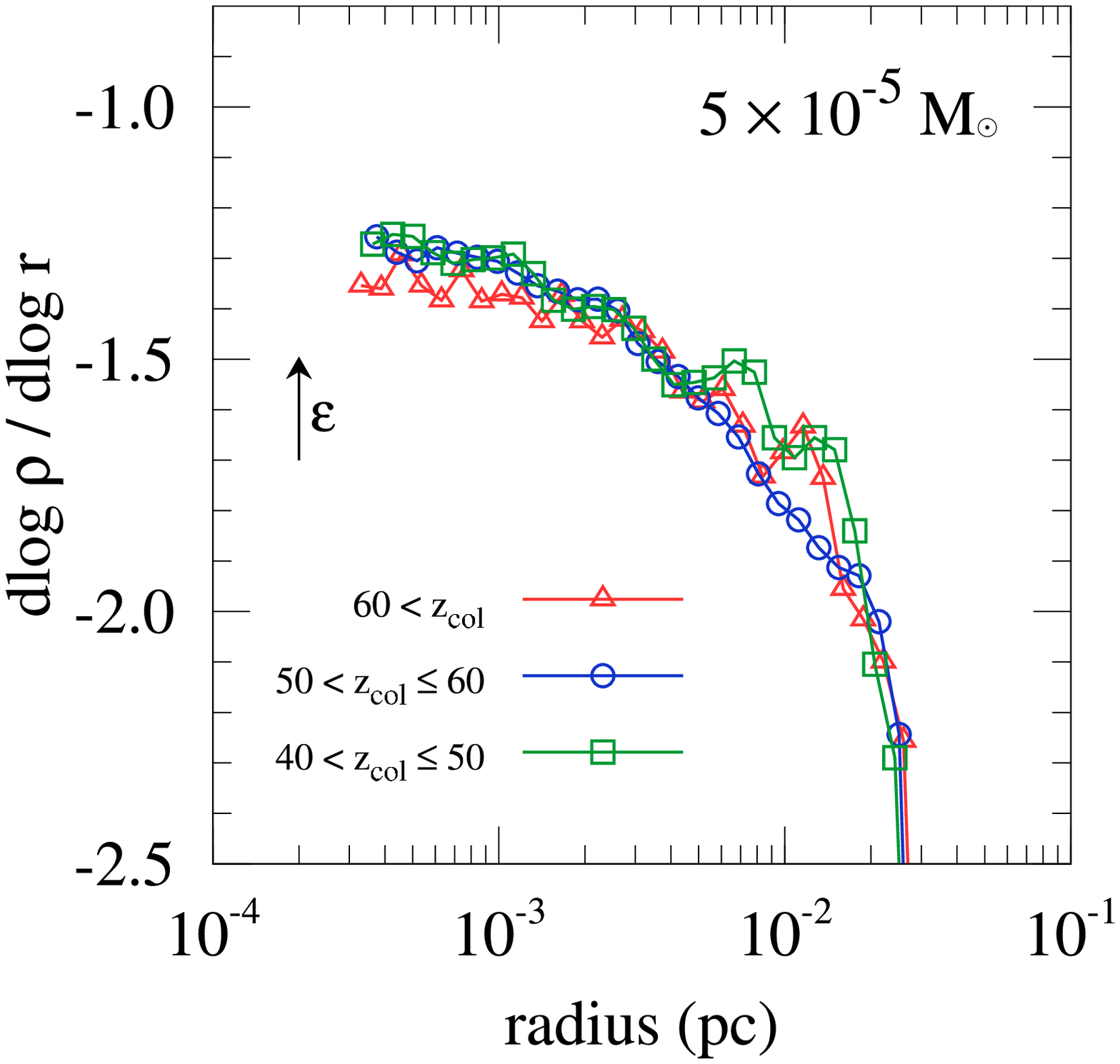}
\includegraphics[width=5.3cm]{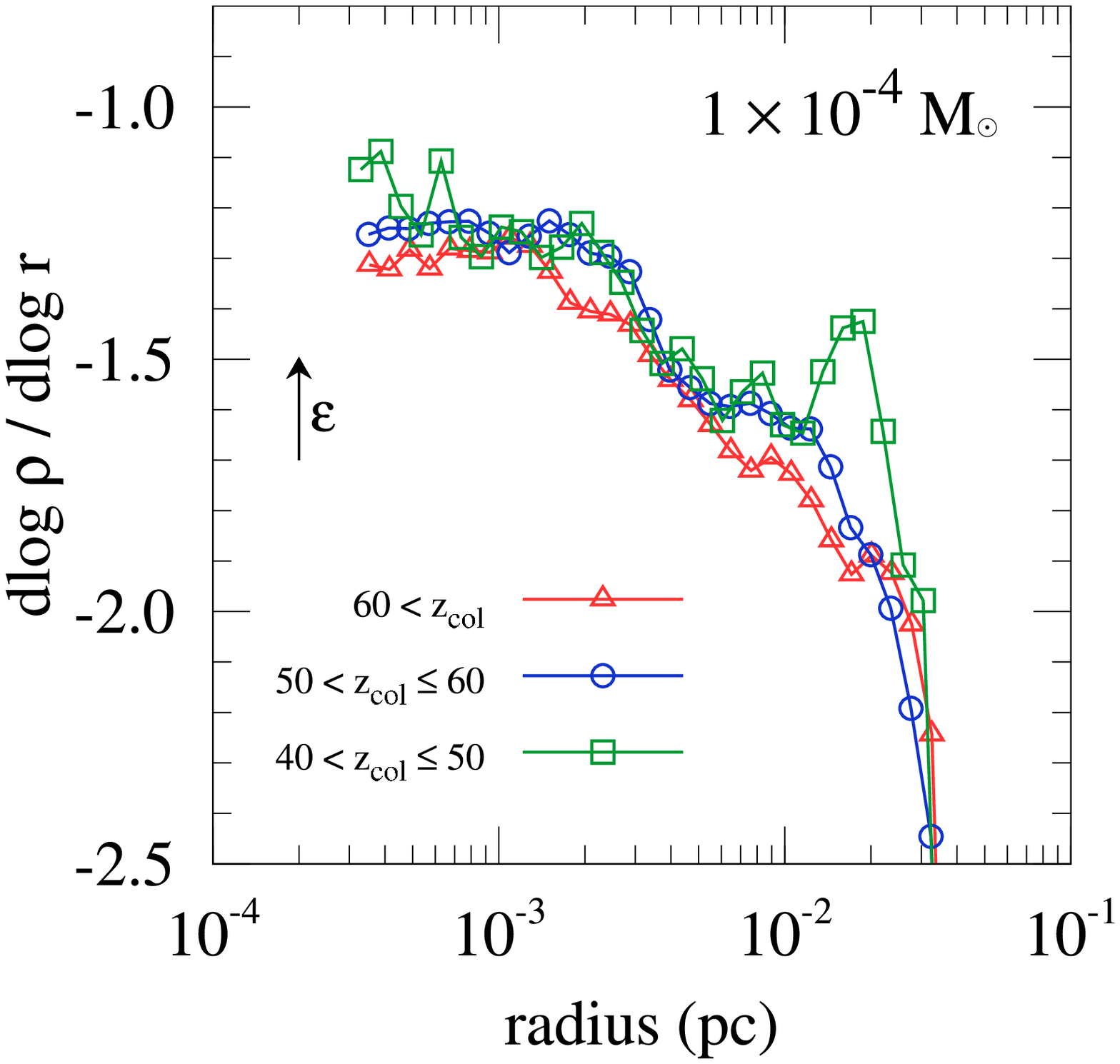}
\caption{ 
Stacked radial density profiles (top panel) and slopes of
profiles (bottom panel) at $z=32$ in the \Af\ simulation.
In each mass bin, the group of halos is categorized into several sub-groups 
according to the collapse epoch defined in the literature.
Arrows indicate the softening length $\varepsilon$.
}
\label{fig:prof_zform}
\end{figure*}

In this subsection, we consider how the density profiles depend on the
halo formation history.  The halo formation epoch is conventionally
defined as the time, at which the most massive progenitor of a halo
gains half of its final mass.  This definition is invalid for halos
near the free streaming scale, since they are the smallest halos, do
not have progenitors in principle.  Therefore, we indicate the halo
formation time by the halo collapse epoch $z_{\rm col}$, the time at
which a halo reaches threshold mass.  We specify the threshold mass as
$1.0 \times 10^{-6} M_{\odot}$, corresponding to 30,000 particles in
simulation \Af.  The masses of the progenitors were calculated as
follows.  First, we collected the unique IDs of all particles
contained in each halo.  We then identified these particles from their
IDs in snapshots of redshifts higher than $z=32$.  For each particle
set, we applied the spherically overdensity method and determined the
mass of the progenitor.

To visualize how the collapse epoch influences the density profile,
we further categorized the halo samples in each mass bin into four
ranges of collapse epoch, $60 < z_{\rm col}$,
$50 < z_{\rm col} \le 60$, and $40 < z_{\rm col} \le 50 $, $z_{\rm col} <
40$.  
Figure \ref{fig:prof_zform} shows the stacked density profiles and slopes at $z=32$,
 were recalculated in each collapse epoch bin for halos in three different mass bins.
Only groups containing more
than 10 halos are shown.  The smallest radii plotted in Figure
\ref{fig:prof_zform} are the reliability limits based on the criteria
proposed by \citet{Power2003}.  Densities inside of these limits are
not plotted.

We can see that halos formed earlier are more
concentrated than those formed later, 
and their central density is higher. 
This figure reinforces the fact that a halo concentration
reflects the cosmic density at its formation time
\citep[e.g.,][]{Bullock2001}.

The slopes of the density profiles appear to slightly depend on the
collapse epoch.  Regardless of their mass, halos with the highest
collapse epochs have slightly steeper cusps than other halos,
particularly at the inner regions.  The inner profiles of halos with
second and third highest collapse epochs are similar although their
central densities differ.  However, since the halos with the largest
collapse epoch are rare at all halo masses, we cannot reliably infer
that that the slope depends on the collapse epoch.  Nonetheless, we
can state that the slope and the collapse epoch are not strongly
correlated, therefore, the halo collapse epoch does not largely
determine the shape of the density profile.

%%%%%%%%%%%%%%%%%%%%%%%%%%%%%%%%%%%%%%%%%%%%%%%%%%%%%%%%%%%%%%%%%%%%%%%%%%%%%%
%%%%%%%%%%%%%%%%%%%%%%%%%%%%%%%%%%%%%%%%%%%%%%%%%%%%%%%%%%%%%%%%%%%%%%%%%%%%%%
\subsection{Origin of Density Profiles}\label{sec:profile_evo}

\begin{figure*}
\centering 
\includegraphics[width=5.3cm]{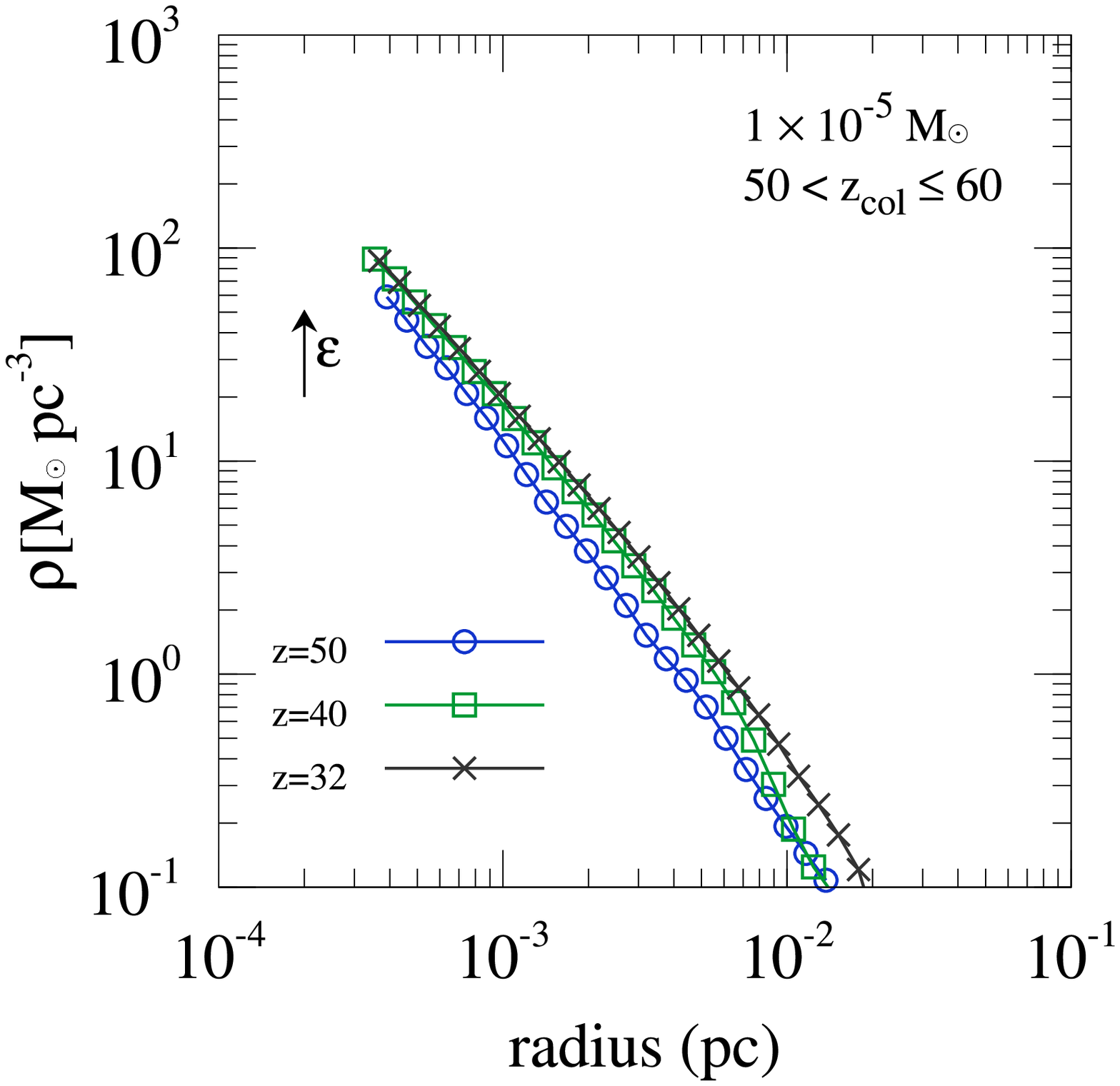}
\includegraphics[width=5.3cm]{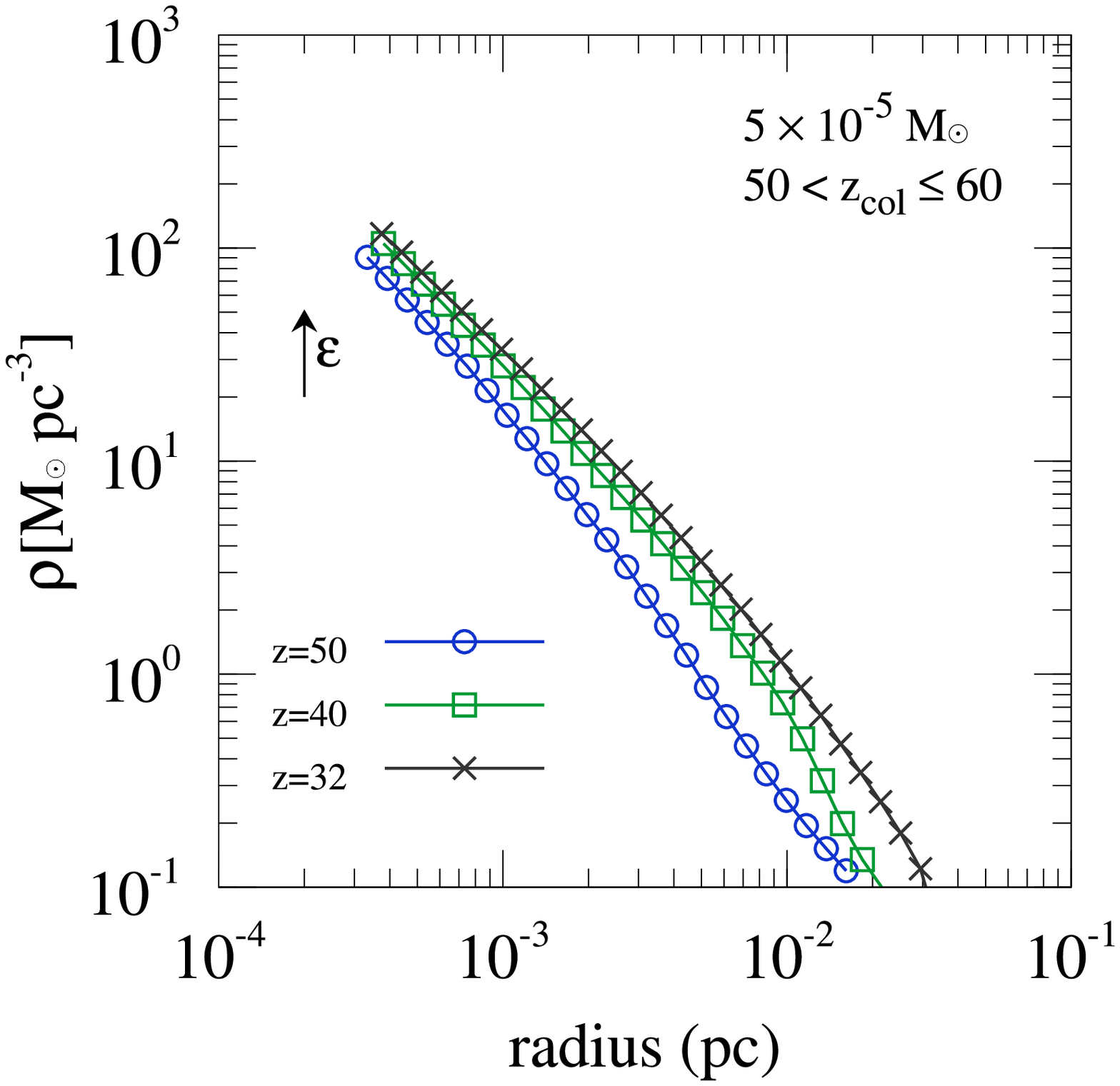}
\includegraphics[width=5.3cm]{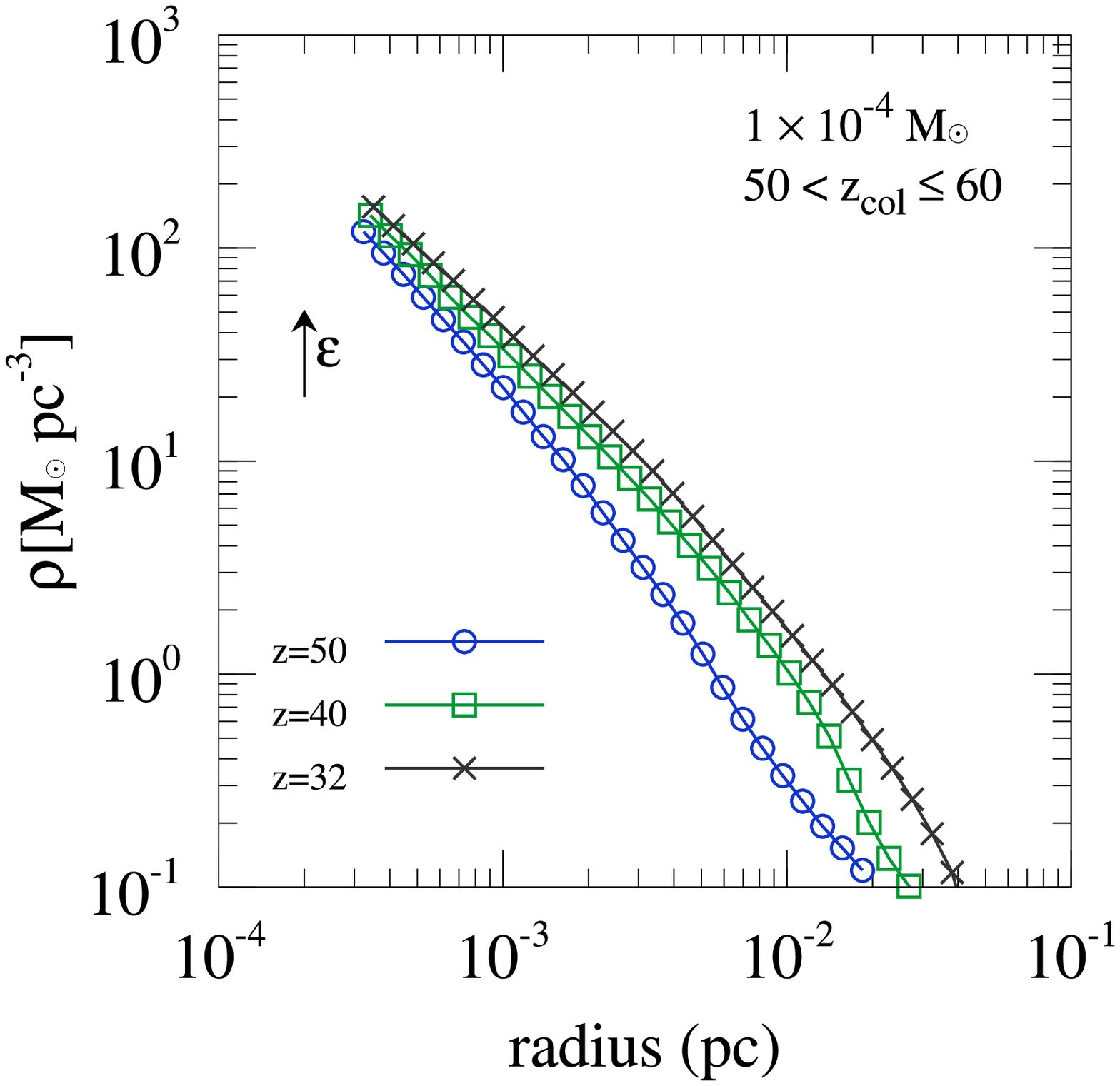}
\includegraphics[width=5.3cm]{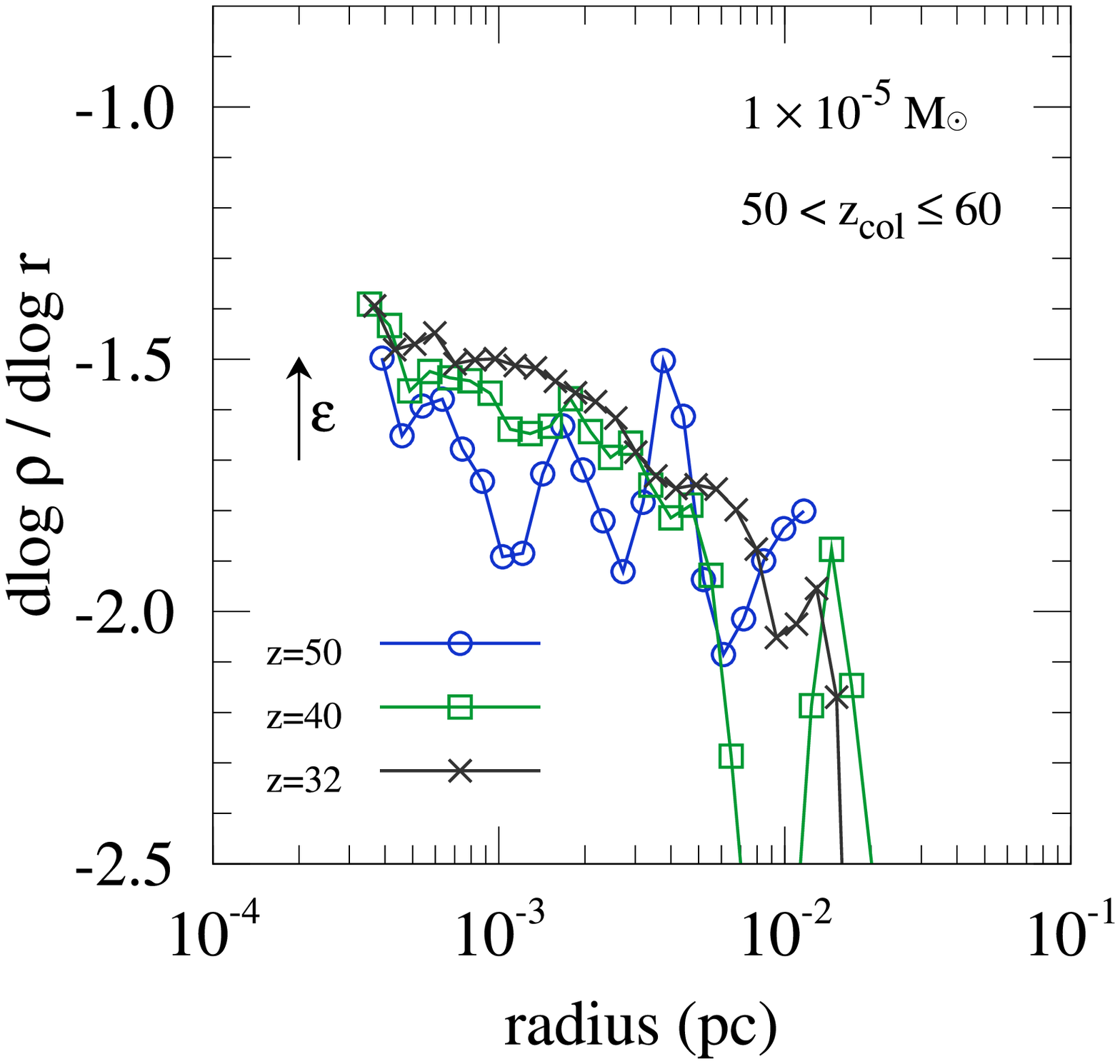}
\includegraphics[width=5.3cm]{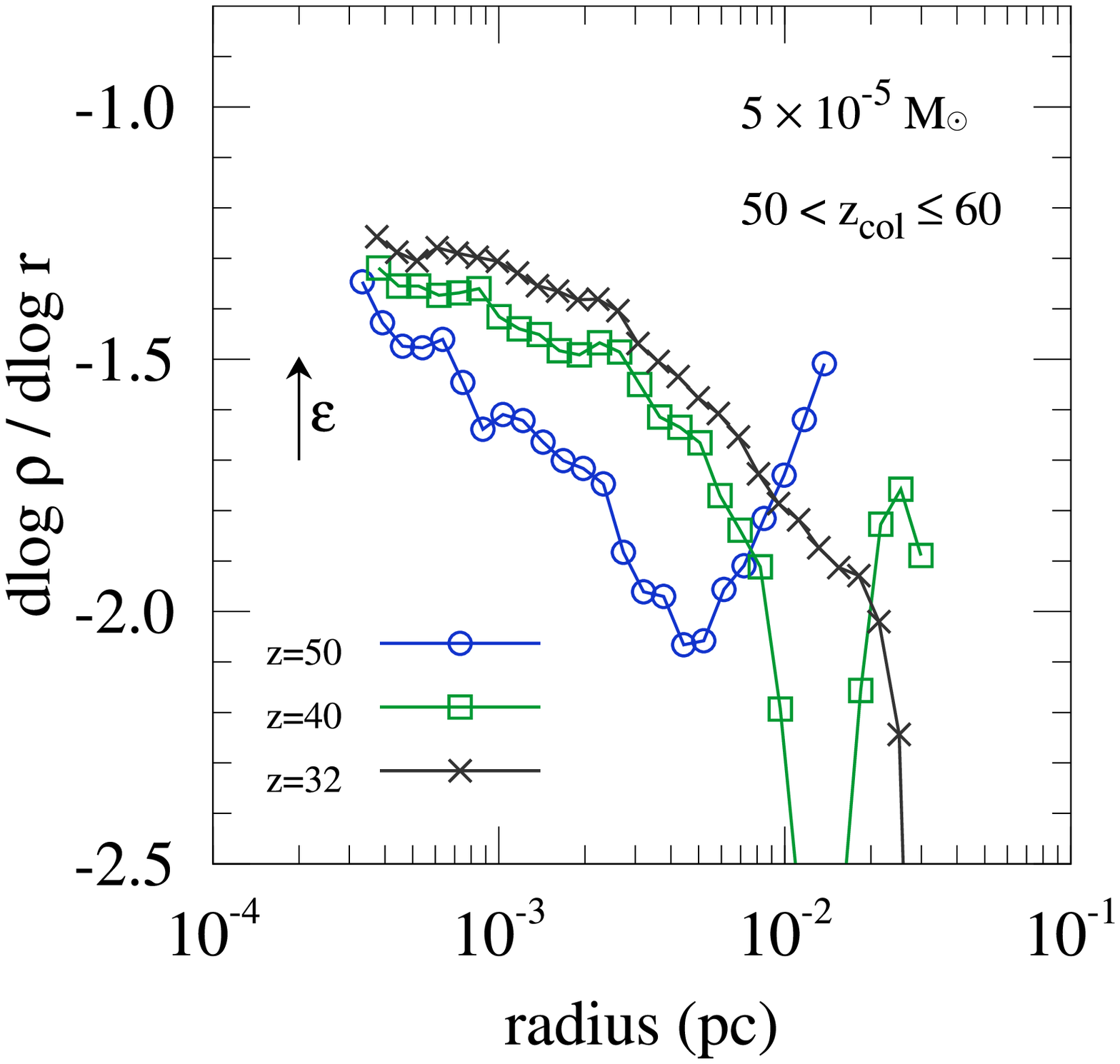}
\includegraphics[width=5.3cm]{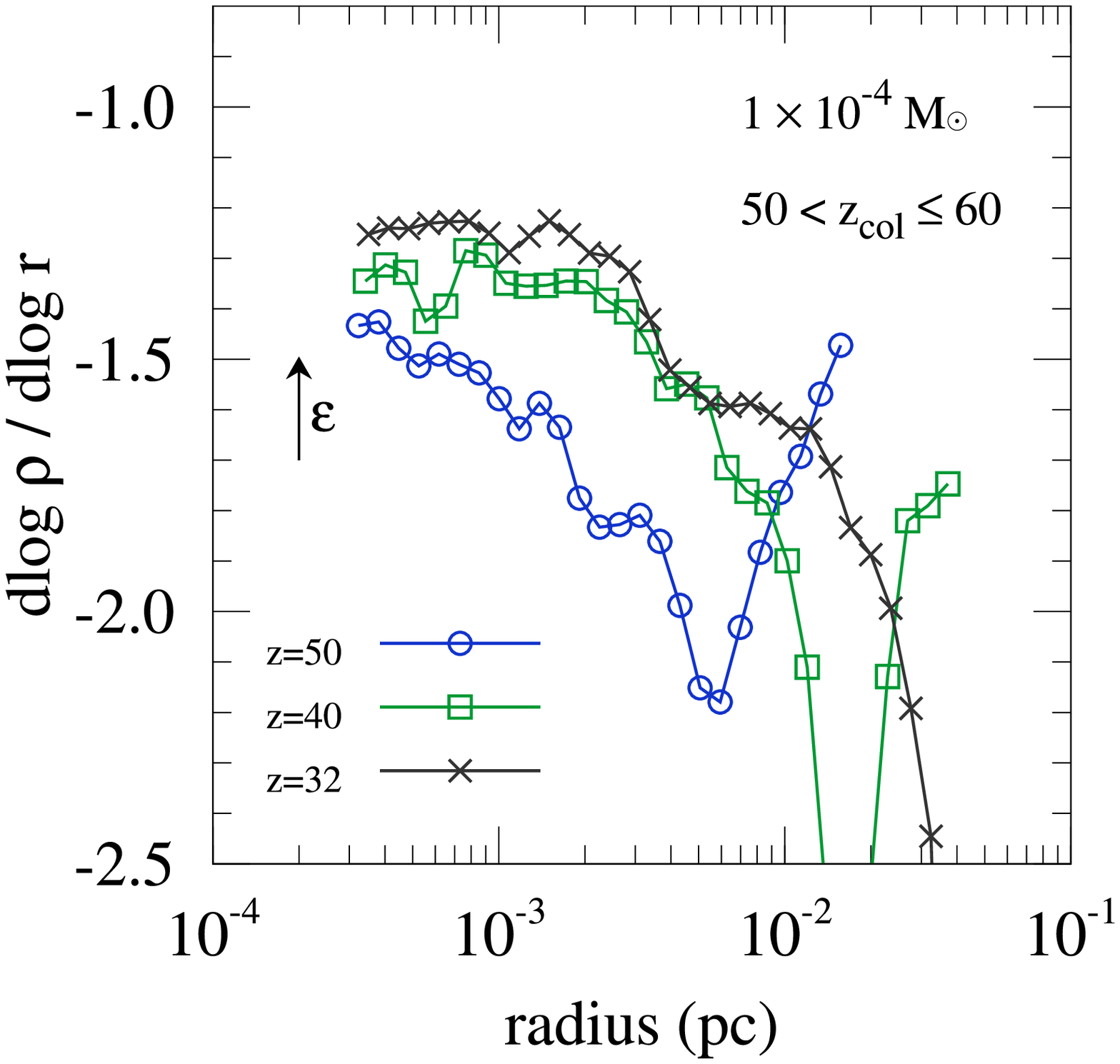}
\includegraphics[width=5.3cm]{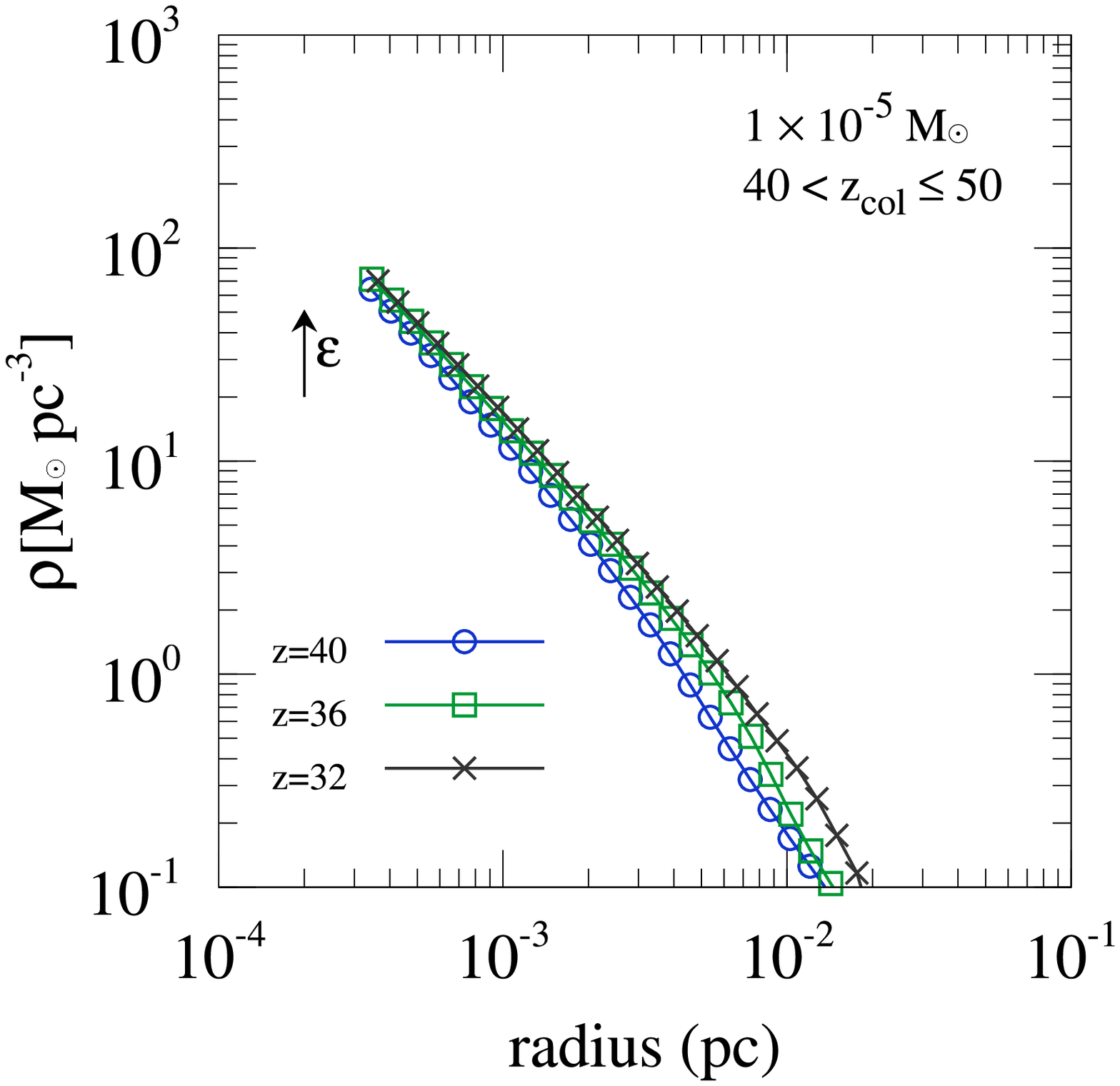}
\includegraphics[width=5.3cm]{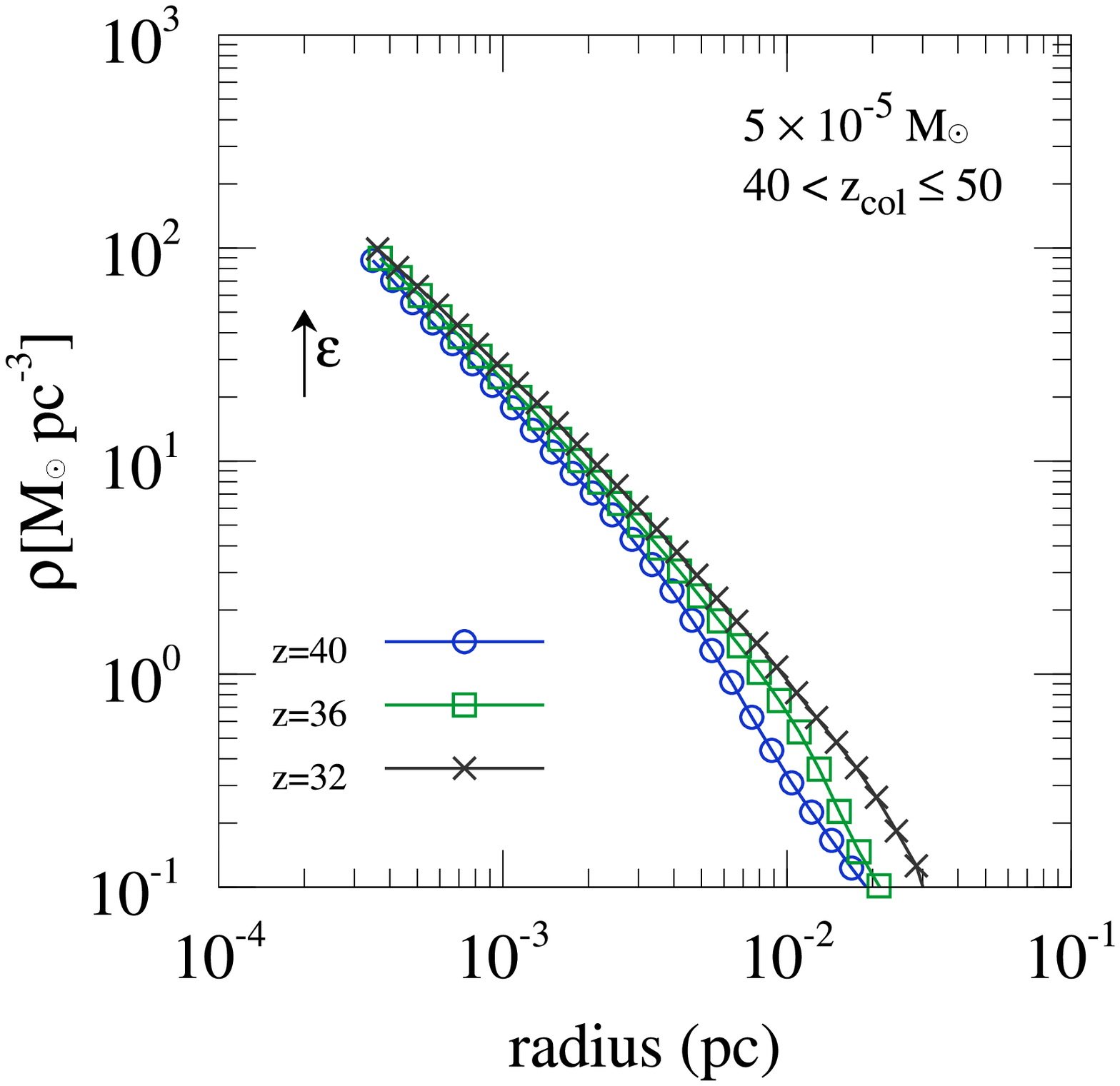}
\includegraphics[width=5.3cm]{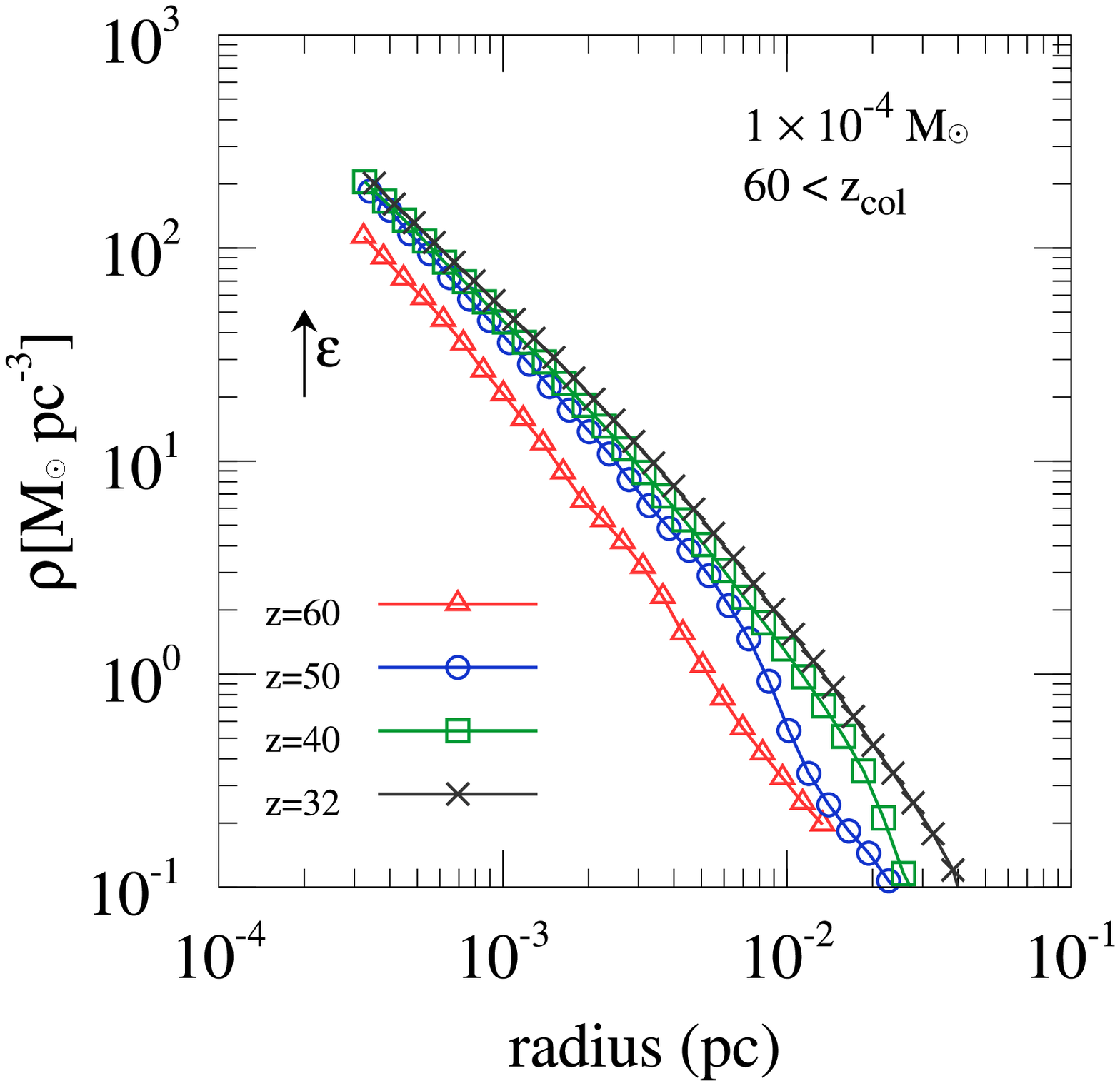}
\includegraphics[width=5.3cm]{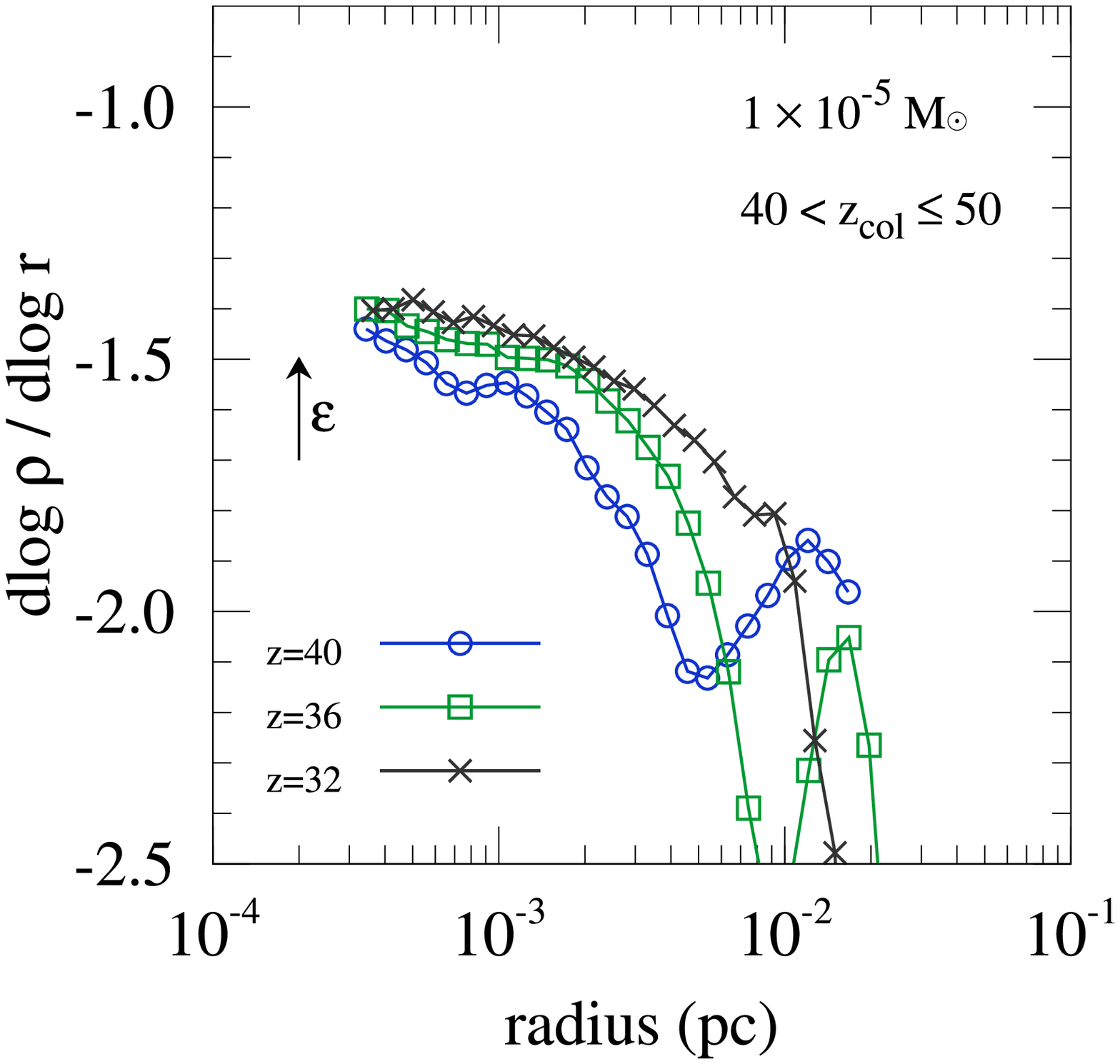}
\includegraphics[width=5.3cm]{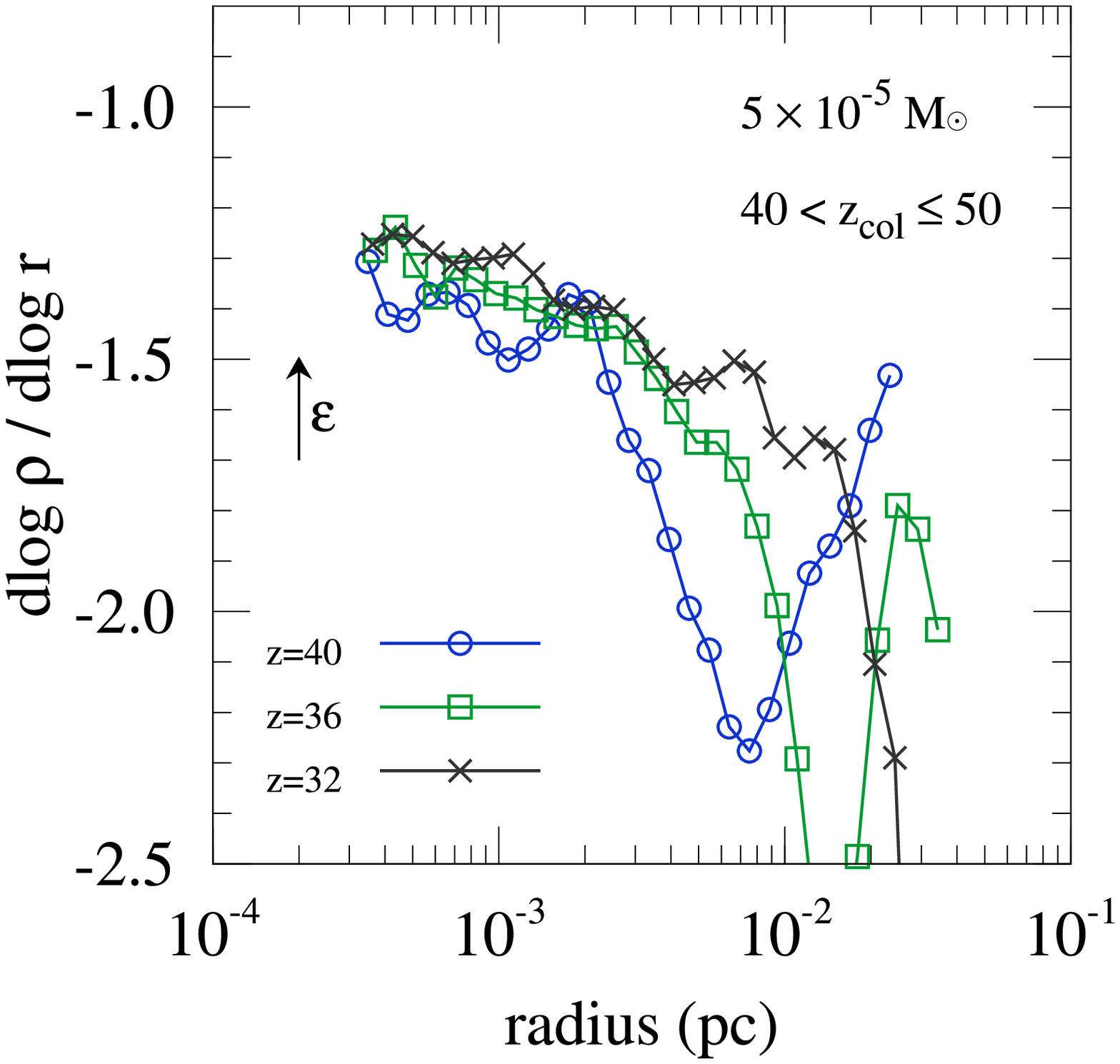}
\includegraphics[width=5.3cm]{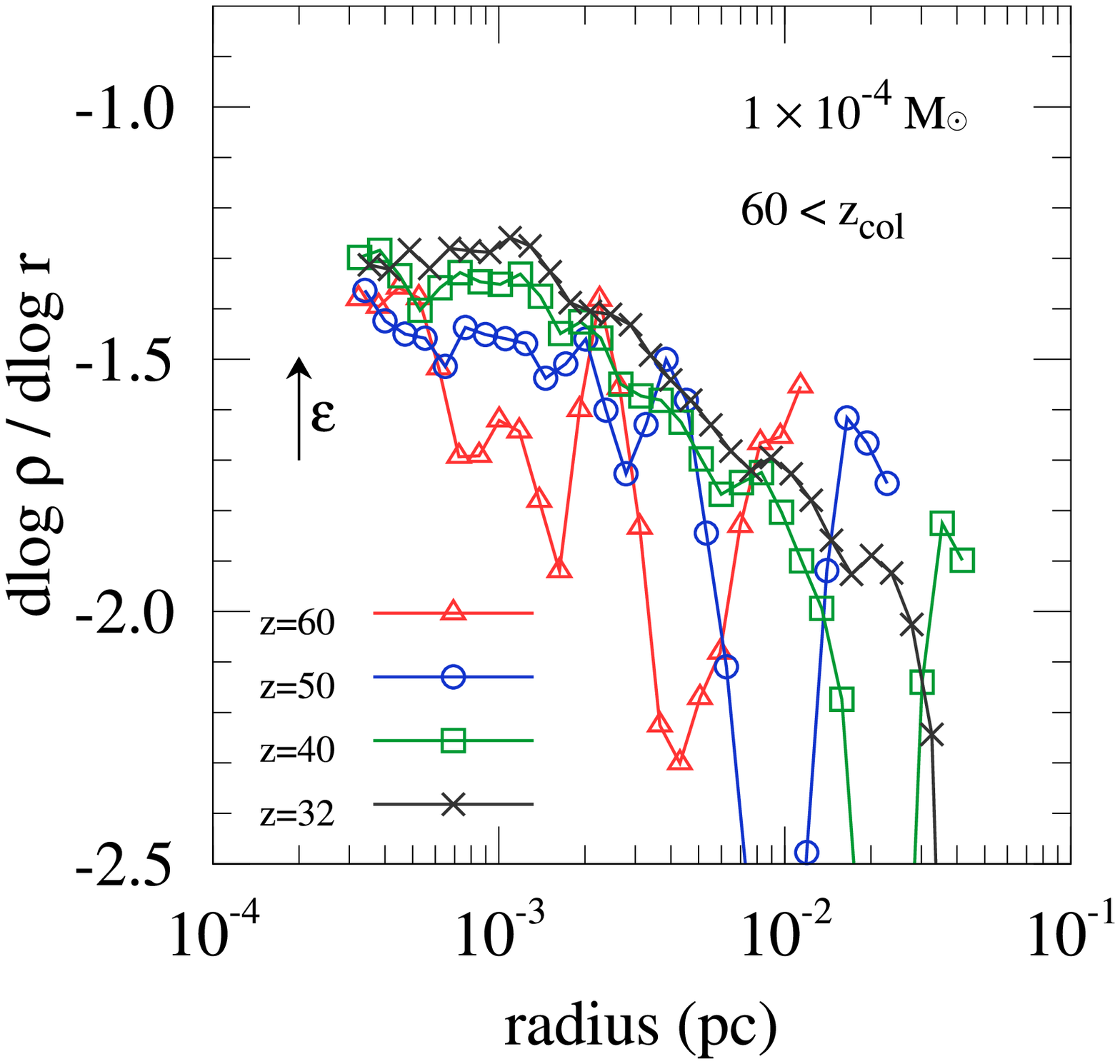}
\caption{ 
Evolution of the stacked radial density profiles (top panel) and slopes of
profiles (bottom panel) in the \Af\ simulation.
Arrows indicate the softening length $\varepsilon$.
}
\label{fig:prof_zform_evo}
\end{figure*}

No confident physical explanation exists for the formation of
structures like the NFW profile found in cosmological simulations.
The physical process that make steeper cusps in the smallest
microhalos is also unclear.  The density structures of the smallest
microhalos probably deviate from the NFW because their formation
processes differ from those of larger halos.  In principle, the
smallest halos contain no subhalos, and are not hierarchically
structured from smaller constituents.  Larger halos form through
repeated mergers of smaller halos.  We sought physical
interpretation of the steep cusps in the smallest halos, but the
solution eluded us at present.  Nevertheless, it is worthwhile to seek
the physical process that the slopes of inner cusps gradually
shallow as the halo accumulates mass.

We focus on the evolution of density profiles of halos, using the samples
introduced in \S \ref{sec:profile3}.  Figure \ref{fig:prof_zform_evo}
shows the evolution of six stacked density profiles with different
collapse epochs and masses.  In all samples, the cusps are clearly steeper
shortly after the collapse epochs than at $z=32$.
The central slopes at the collapse are approximately $-1.5$, 
comparable to those of the smallest microhalos in
\citet{Ishiyama2010}, Figures \ref{fig:prof_comp}, \ref{fig:prof}, 
\ref{fig:m-alpha}, and \ref{fig:m-alpha_bin}.  
Both central and outer
densities
gradually increase with decreasing redshift.
The central densities grow rather more moderately as than the 
outer densities.

In halos with $50 < z_{\rm col} \le 60$, the central densities and
slopes have almost ceased growing before $z=40$.  From $z=50$ to
40, the central slopes dramatically evolve. Differences among the
profiles in each mass bin are already evident at $z=40$.  The cusps of
halos with higher final mass are shallower than those of lower mass
halos.  From $z=40$ to 32, the central densities and slopes appear
to approach constant values.  On the other hand, the outer densities
continue to grow beyond $z=40$, as external dark matter smoothly
accretes in the outer halo regions.

\begin{figure*}
\centering 
\includegraphics[width=5.3cm]{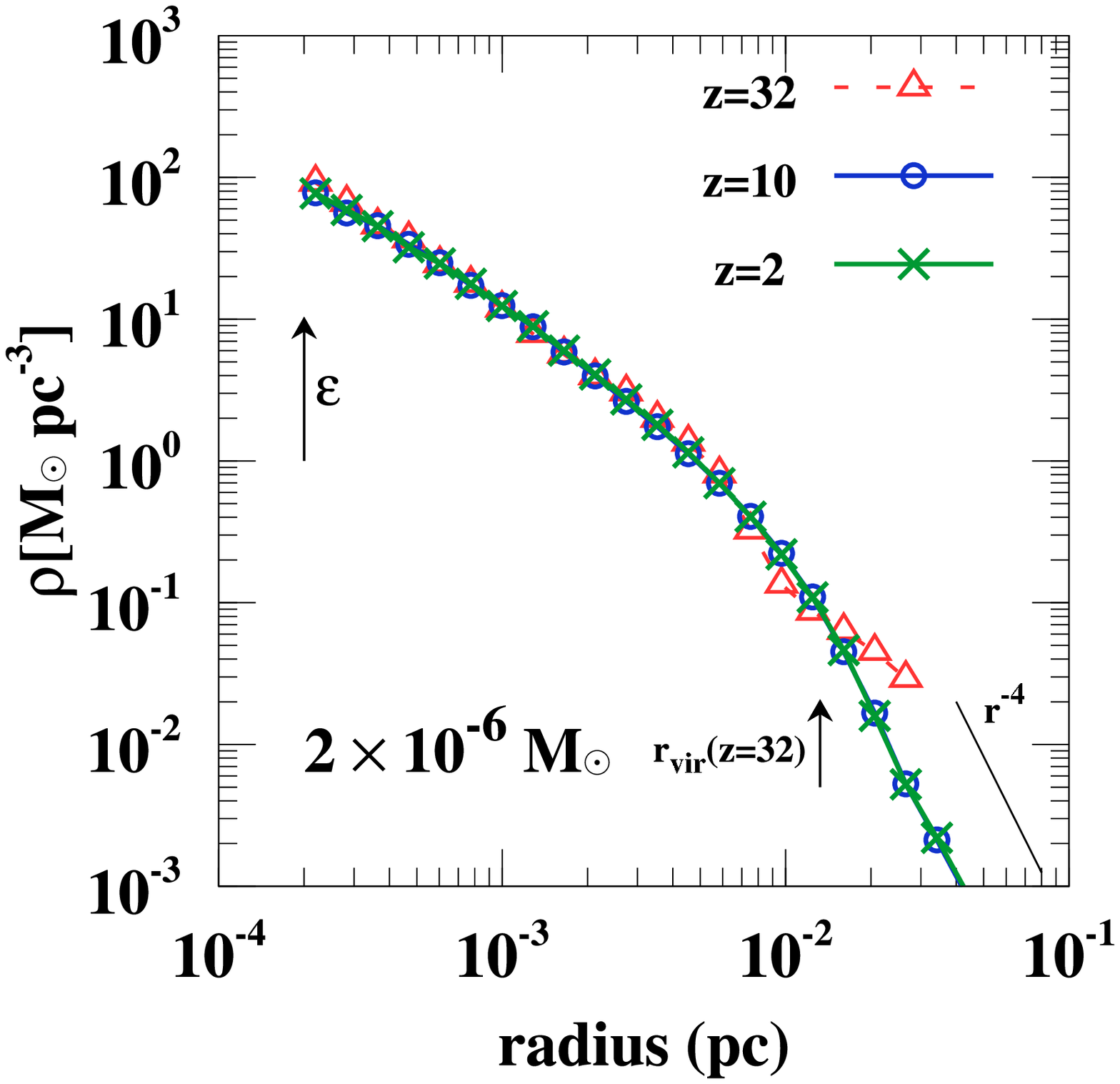}
\includegraphics[width=5.3cm]{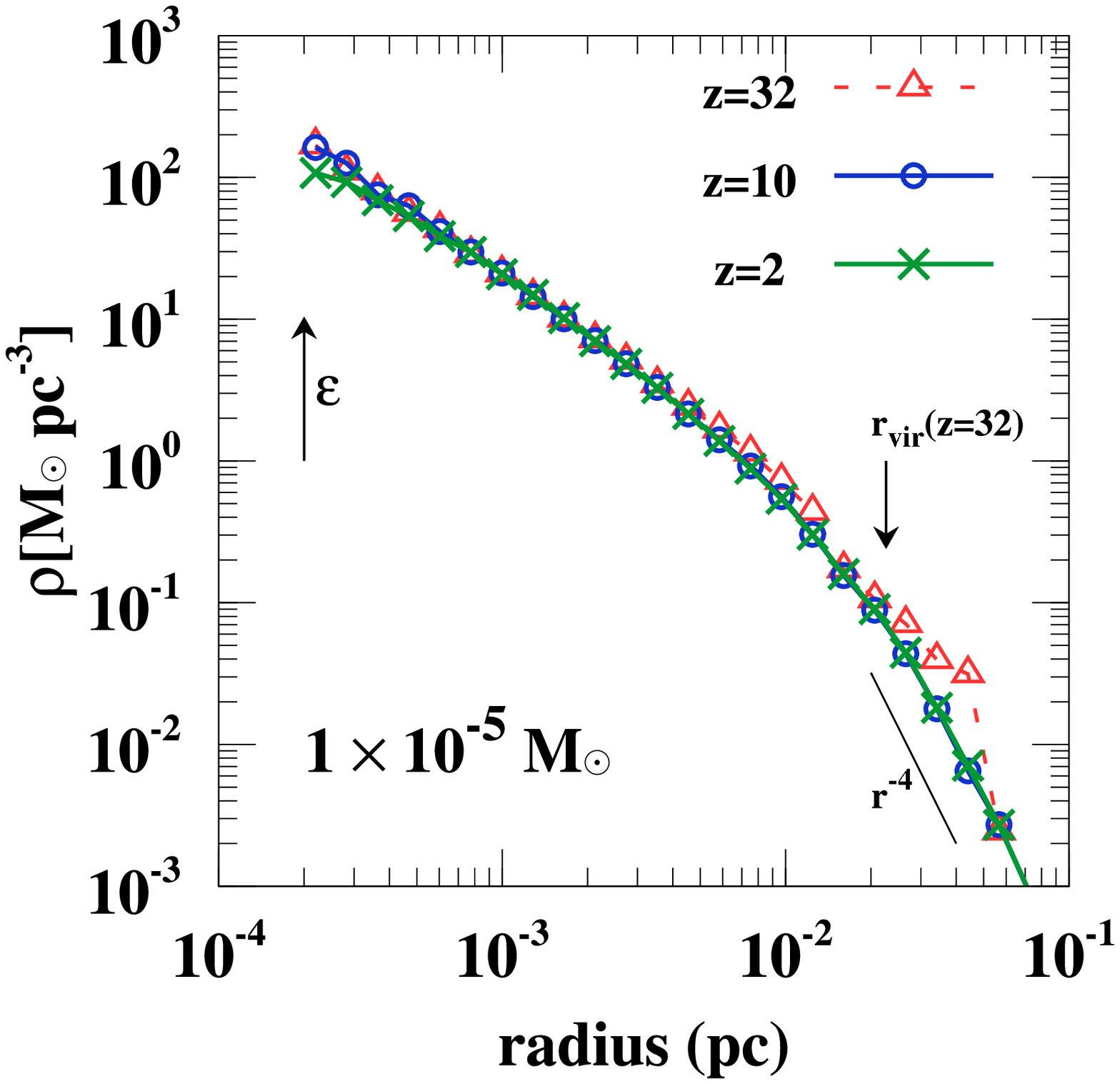}
\includegraphics[width=5.3cm]{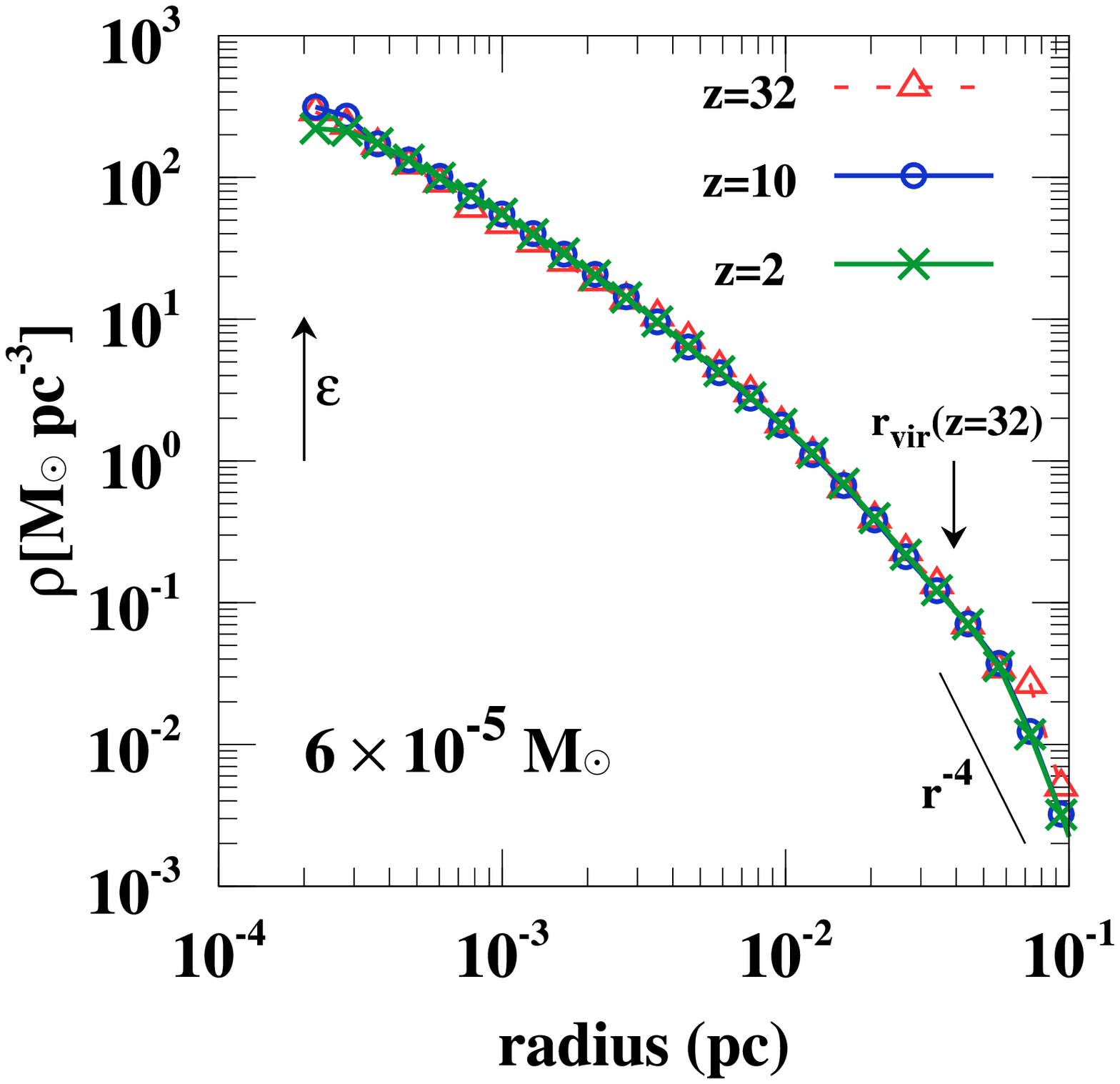}
\caption{ 
Evolution of density profiles toward low redshifts. 
Arrows indicate the softening length $\epsilon$ and 
the halo virial radius at $z=32$, $r_{\rm vir}(z=32)$. 
}
\label{fig:prof_evo}
\end{figure*}

Such trends also emerge in halos with $40 < z_{\rm col} \le 50$.
Although halos with $60 < z_{\rm col}$ display the clearest trends,
the number of halos in this category is only sufficient for $1.0
\times 10^{-4} M_{\odot}$ halos.  These figures indicate a non
self-similar evolution of the density structure. By contrast,
galaxy-sized and cluster-sized halos develop in a self-similar way
\citep[e.g.,][]{Fukushige2001}.  In principle, the fraction of major
mergers in halos near the free streaming scale is larger than that in
larger halos because of the absence of smaller subhalos.  
On the other hand,
major merger simulations of two isolated halos showed that the steeper cusp
is preserved \citep[e.g.,][]{Boylan2004, Zemp2008}, consistent with a
theoretical model \citep{Dehnen2005}, or slightly reduced inner
slope emerges \citep[e.g.,][]{Boylan2004}.  Therefore, repeated and
nearly simultaneous major mergers would majorly contribute to
shallowing the cusp slope in halos near the free streaming scale.

%%%%%%%%%%%%%%%%%%%%%%%%%%%%%%%%%%%%%%%%%%%%%%%%%%%%%%%%%%%%%%%%%%%%%%%%%%%%%%
%%%%%%%%%%%%%%%%%%%%%%%%%%%%%%%%%%%%%%%%%%%%%%%%%%%%%%%%%%%%%%%%%%%%%%%%%%%%%%
\subsection{Stability of the Density Profile}

One may ask whether the density structures in the simulations of
the cutoff model have not stabilized.  Are these density profiles with
steep cusps transient?  Do they transit to shallower
cusps as seen in the profile of the \Bf\ simulation?  To address
these questions, we performed additional simulations to follow the
evolution of these density structures.

We randomly selected three halos at $z=32$ and evolved them as
isolated systems.  We extracted a $2 \times 10^{-6} M_{\odot}$ halo
from then \As\ simulation, and two halos of masses $1 \times 10^{-5}
M_{\odot}$ and $6 \times 10^{-5} M_{\odot}$ from the \Af\ simulation.
We identified particles within $2r_{\rm vir}$ of the potential minima
of these halos and re-simulated the evolution using only the tree part
of GreeM code \citep{Ishiyama2009b, Ishiyama2012} from $z=32$ to 2.
The corresponding time interval is about 3.2 Gyr.  In all simulations,
the constant softening value $2 \times 10^{-4} \rm pc$ was used and
the opening angle for the tree method was 0.5.  The properties of
these halos are summarized in Table \ref{tab2}.

\begin{table*}[t]
\centering
\caption{Details ofAadditional Simulations
Here, $N(<r_{\rm vir})$, $N(<2r_{\rm vir})$, $m$, $M_{\rm vir}$, and $r_{\rm vir}$ 
are the number of particles within $r_{\rm vir}$ and $2r_{\rm vir}$, 
the mass resolution, 
the virial mass, and the virial radius of each halo, respectively.
The last column indicates the base simulation. 
In all simulations, the gravitational softening length 
was $2 \times 10^{-4} \rm pc$.
}\label{tab2}
\begin{tabular}{cccccc}
\hline\hline
$N(<r_{\rm vir})$ & $N(<2r_{\rm vir})$ & $m (M_{\odot})$ & $M_{\rm vir} (M_{\odot})$ & $r_{\rm vir}(\rm pc)$ & simulation\\
\hline
532713  & 1128383 & $4.3 \times 10^{-12}$ & $2.29 \times 10^{-6}$ & $1.32 \times 10^{-2}$ & \As \\
339918  & 694313  & $3.4 \times 10^{-11}$ & $1.16 \times 10^{-5}$ & $2.26 \times 10^{-2}$ & \Af \\
1737815 & 3239747 & $3.4 \times 10^{-11}$ & $5.94 \times 10^{-5}$ & $3.92 \times 10^{-2}$ & \Af \\
\hline 
\end{tabular}
\end{table*}

Figure \ref{fig:prof_evo} plots the evolution of density structures of
these halos.  The density structures are clearly stable from a few
$\epsilon$ to $r_{\rm vir}$.  Beyond $r_{\rm vir}$, the outskirts of
all halos expand and the density profile appears to follow $\rho(r)
\propto r^{-4}$, because the outskirt of each original halo was
confined within $2r_{\rm vir}$.  When a halo suddenly loses its
outskirt, it experiences strong gravitational disturbance and
establishes a new equilibrium state.  This situation naturally leads
to $\rho(r) \propto r^{-4}$ \citep[e.g.,][]{Jaffe1987, Makino1990}.
However, this scenario may be unrealistic in the cosmological context, 
since continuous infall of matter and 
intermittent merger processes occur, 
which would increase the densities in outer regions.

Around $\epsilon$, slight density decreases are observed at $z=2$. These 
decreases are obviously numerical artifacts introduced by two-body 
relaxation.  Therefore, the density structures with steep cusps appear 
to be stable rather than transient, and exist in an equilibrium state.

From these results, we can infer how these halos born in the early
universe exist at present. If these halos remain sufficiently isolated
from larger systems, they should wander throughout the universe
while retaining their steep cusps.  If 
captured by larger systems and exist as subhalos, 
their densities will be altered by 
tides from the larger systems.
Since tidal mass loss mainly occurs in the outer regions \citep{Diemand2007b}, 
the high central density of these halos would remain
after captured by the larger system.

Finally, what subhalo mass function can we observe in a halo as
massive as the Milky Way halo?  In high resolution simulations of such
halos, the mass function appeared to scale as $dn/dm \propto
m^{-(2\mbox{\scriptsize --}1.8)}$, although this trend is not
conclusive \citep[e.g.,][]{Berezinsky2003, Ishiyama2009}.  However,
none of previous simulations could resolve the free streaming scale,
since earth mass halos are smaller than Milky Way mass halos by 18
orders of magnitude.  Subhalos resolved in previous simulations have
similar density structures to those of their host halo.  Since these
subhalos are much larger than the
CDM cutoff scale, 
single slope in
their mass function is expected.  However, whether similar mass functions 
apply to halos near the cutoff scale
is unclear, since the cusp
in halos near the free streaming scale are steeper than those in
larger halos.

In the \Af\ simulation, more than thirty halos contained than 10
million particles at $z=32$, sufficient for analyzing the mass
function and the spatial distribution of subhalos.  The statistics of
these subhalos will be presented elsewhere.

%%%%%%%%%%%%%%%%%%%%%%%%%%%%%%%%%%%%%%%%%%%%%%%%%%%%%%%%%%%%%%%%%%%%%%%%%%%%%%
%%%%%%%%%%%%%%%%%%%%%%%%%%%%%%%%%%%%%%%%%%%%%%%%%%%%%%%%%%%%%%%%%%%%%%%%%%%%%%
%%%%%%%%%%%%%%%%%%%%%%%%%%%%%%%%%%%%%%%%%%%%%%%%%%%%%%%%%%%%%%%%%%%%%%%%%%%%%%
%%%%%%%%%%%%%%%%%%%%%%%%%%%%%%%%%%%%%%%%%%%%%%%%%%%%%%%%%%%%%%%%%%%%%%%%%%%%%%
\section{Discussions}\label{sec:discussion}

\subsection{Contributions of Halos Near the Free Streaming Scale 
to Gamma-ray Annihilation Signals} 

The gamma-ray luminosity of a halo by neutralino self-annihilation
seen from a distant observer 
is calculated by the volume integral of
the density squared.
Any subhalos boost the annihilation luminosity of a halo and 
the boost factor $B(M)$ is defined as 
\citep[e.g.,][]{Strigari2007}
\begin{eqnarray}
B(M) = \frac{1}{L(M)} \int^M_{M_{\rm min}}\frac{dn}{dm} \, [1+B(m)] \, L(m) \, dm, 
\label{eq:bf}
\end{eqnarray}
where, $L(M)$ is the annihilation luminosity of a halo of mass $M$ 
without subhalos,
and $dn/dm = A/M (m/M)^{-\zeta}$ is the subhalo mass function.

The boost factor of a Milky Way sized halo has been estimated
by many studies and ranges from a few to several tens
\citep[e.g.,][]{Colafrancesco2006,Diemand2007,Diemand2008,
Kamionkowski2010, Kuhlen2012, Sanchez-Conde2013}.
Some works give several hundreds \citep{Springel2008b}.

In many early studies, annihilation luminosities and boost factors, have
been evaluated assuming that halos near the free streaming scale
follow the NFW profile.  The steeper cusps obtained in our simulations
could significantly enhance the signals.  In this section, we evaluate
boost factors using the density profiles obtained in our simulations.
However, many uncertainties, the evolution of these halos in the Milky
Way, subhalo mass function, and the profile of the Milky Way halo,
weaken the reliability of estimation.  Nevertheless, such estimation
is worthy to attempt and gives us some insights about the boost
factor.

As models of halos near the free streaming scale,
we consider three models.
\begin{enumerate}
\item {\it \Mf}.
The profile of Equation (\ref{eq:doublepower})
  with the mass--shape relation of Equation (\ref{eq:fita}), and the
  constant concentration $c=1.35$ at $z=32$, consistent with our
  cutoff simulation.
\item {\it \Ms}. 
The profile of Equation (\ref{eq:doublepower}) with the
mass--shape relation of Equation (\ref{eq:fita}), and 
the mass--concentration relation proposed by \citet{Sanchez-Conde2013}. 
This NFW concentration is converted to that of 
Equation (\ref{eq:doublepower}) by the way described in 
\S \ref{sec:concentration}.
\item {\it \Mt}. 
The NFW profile and the mass--concentration relation proposed by 
\citet{Sanchez-Conde2013}. 
\end{enumerate}
In the mass--shape relation of Equation (\ref{eq:fita}), the inner
shape becomes $\alpha=1$ at the mass $\sim 5.6 \times 10^{-3}
M_{\odot}$.  We assume that the density profile is identical to the
NFW profile above this mass scale.  We used the subhalo mass function
of $\zeta=1.9$ and $2.0$, corresponding normalization factors are
$A=0.03$ and $0.012$ \citep[e.g.,][]{Sanchez-Conde2013} and calculated
the annihilation luminosity of each mass halo by numerically integrating
the square of Equation (\ref{eq:doublepower}) within the range
$10^{-5}$ pc to $r_{\rm vir}$, with the parameters of adopted model.
Then the boost factor was calculated by numerically integrating
Equation (\ref{eq:bf}) from $M_{\rm min}=10^{-6} M_{\odot}$ under the
condition $B(M_{\rm min})=0$.

\begin{figure}
\centering 
\includegraphics[width=8.5cm]{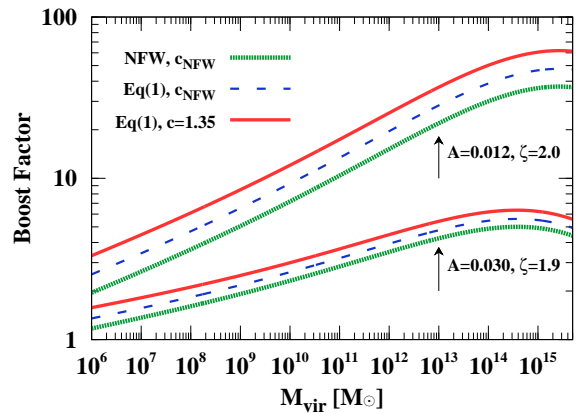}
\caption{ 
Annihilation boost factor as a function of the halo mass.
The results of three models of halos near 
the free streaming scale are shown.
The subhalo mass function $dn/dm = 0.012/M (m/M)^{-2.0}$
gives upper three curves.
The subhalo mass function $dn/dm = 0.030/M (m/M)^{-1.9}$
gives lower three curves.
}
\label{fig:bf}
\end{figure}

Figure \ref{fig:bf} shows the annihilation boost factor as a function
of the halo mass.  The result of \Mt\ model is identical to that shown
in \citet{Sanchez-Conde2013}.  The models with steeper inner (\Mf\ and
\Ms) raise the boost factor moderately.  The model \Mf\ gives larger
enhancement than the model \Ms, since its concentrations are larger 
as seen in Figure \ref{fig:m-c_fit}.  The boost factors of a Milky Way
sized halo ($M=2.0 \times 10^{12} M_{\odot}$) with the
$\zeta=2.0$ subhalo mass function are $\sim 17, 22$ and $29$ for
models of \Mt, \Ms\ and \Mf, respectively.  Those with the
$\zeta=1.9$ subhalo mass function are $\sim 3.7, 4.2$ and $4.8$.
Strongly depending on the subhalo mass function and the adopted
concentration model, the steeper inner cusps of halos 
near the free streaming scale
enhance the annihilation luminosity of a Milky Way sized halo between
12\% to 67\%.

When simple single power law mass--concentration relations are
extrapolated to microhalo scales, the resulting boost factor is
several hundreds for a Milky Way sized halo
\citep[e.g.,][]{Springel2008b}, and $\sim 1000$ for a cluster sized
halo \citep{Pinzke2011, Gao2012}.  Such very large boost factors are
clearly ruled out as seen in Figure \ref{fig:bf}, since our high
resolution simulations rule out single power law mass--concentration
relations.

The scatter in the density profiles seen in Figures
\ref{fig:m-alpha_bin} and \ref{fig:m-c_bin} could influence the boost
factor.  To evaluate the effect of the scatter, we calculated the
annihilation luminosity of each simulated halo by numerically
integrating the square of Equation (\ref{eq:doublepower}) , with the
parameters obtained in the fitting of each halo.  We find that the
average luminosity is 28\% smaller than that we do not consider the
scatter.  Thus, the scatter in the density profiles would not cause
large effect on the boost factors.

Our work has improved the accuracy of the estimation of the boost
factor.  However, some uncertainties still exist.  It is unclear
whether the subhalo mass function can be extrapolated to the scale of
microhalos, since the central density cusp is steeper in these halos
than in larger halos.  The boost factor strongly depends on the cutoff
mass scale \citep[e.g.,][]{Anderhalden2013, Sanchez-Conde2013}.
Further studies about the evolution of microhalos in larger halos and
a cutoff scale independent model of density profiles, are
required to estimate the boost factor more robustly.

\subsection{Analogy with Warm Dark Matter Simulations}
Our studies are closely related to warm dark matter simulations.
Although mass scales are largely different, the mechanism to impose
the cutoff in the matter power spectrum is similar.  Thus, the
structure of halos near the cutoff scale of warm dark matter
particles should be similar to those obtained in our simulations.

The mass--shape relation of Equation (\ref{eq:fita}) 
predicts that the slopes with the cutoff and no cutoff models 
are similar at around $10^{-3}M_{\odot}$, 
which is about three orders larger than the cutoff scale.   
This prediction is consistent with recent warm dark matter simulations
\citep[e.g.,][]{Anderhalden2012, Lovell2013}
that gave density profiles similar to the NFW profile 
for the Milky Way mass halos, 
which are larger than the cutoff scale of warm dark matter particles
by three orders of magnitude. 

Our results indicate that the cusps of warm dark matter halos 
below the Milky Way mass would be steeper than that of the NFW profile.
However, there is an absence of simulation data. 
Our understanding of halo structures will advance by
high resolution warm dark matter simulations, which resolve
halos near its cutoff scale.

%%%%%%%%%%%%%%%%%%%%%%%%%%%%%%%%%%%%%%%%%%%%%%%%%%%%%%%%%%%%%%%%%%%%%%%%%%%%%%
%%%%%%%%%%%%%%%%%%%%%%%%%%%%%%%%%%%%%%%%%%%%%%%%%%%%%%%%%%%%%%%%%%%%%%%%%%%%%%
 %%%%%%%%%%%%%%%%%%%%%%%%%%%%%%%%%%%%%%%%%%%%%%%%%%%%%%%%%%%%%%%%%%%%%%%%%%%%%%
%%%%%%%%%%%%%%%%%%%%%%%%%%%%%%%%%%%%%%%%%%%%%%%%%%%%%%%%%%%%%%%%%%%%%%%%%%%%%%
\section{Summary}\label{sec:summary}

By means of unprecedentedly large cosmological $N$-body simulations,
we studied the structures of dark matter halos near the free streaming
scale over a wide mass range ($\sim 10^{-(6\mbox{\scriptsize --}4)}
M_{\odot}$).  These simulations enable to resolve halos with
sufficient resolution and cover volumes large enough to perform
statistical studies.  Two different initial matter power spectra were
adopted, one accounting for the free streaming damping, the other
ignoring this effect.  We investigated the effect of the free
streaming damping on structures of halos. Our main results are
summarized below.

\begin{enumerate}
\item 
The central cusp of the smallest microhalos scales as 
$\rho \propto r^{-(1.5\mbox{\scriptsize --}1.3)}$, much steeper than that of the
NFW profile, but consistent with previous works
\citep{Ishiyama2010, Anderhalden2013}.  The central cusp
becomes gradually shallower as the halo mass increases.  For halos of mass $5
\times 10^{-5} M_{\odot}$, the central slope is around $-1.3$.

\item 
The density profiles of these halos are not well fitted by the NFW profile, 
but can be fitted to a double power law function 
(Equation (\ref{eq:doublepower})).
Within the mass range of this study, 
the shape parameter $\alpha$ is given as
$\alpha = -0.123 \log( M_{\rm vir} / 10^{-6}M_{\odot}) + 1.461$. 

\item 
The concentration parameter shows little dependence on the halo mass
within the mass range of this study.  The median concentration in the
cutoff and the no cutoff models ranges
$1.2\mbox{\scriptsize --}1.7$ and
$1.8\mbox{\scriptsize --}2.3$, respectively, corresponding to
conventional concentrations (based on the NFW profile at $z=0$) of
$60\mbox{\scriptsize --}70$. These results support suggestions
that the concentration does not depend on mass in a single power law
fashion.

\item
No strong correlation exists between the slope and the collapse epoch,
indicating that halo collapse is not important in determining the
shape of the density profile.  The density profile does not evolve
self-similarly, unlike larger halos such as galaxy-sized and
cluster-sized halos.  Shortly after collapse, 
the cusps are steeper and their profiles are
similar to those of the smallest halos.  The cusp slope in halos
near the free streaming scale should be predominantly reduced by
merger processes.

\item
These density profiles with steep cusps 
are not transient and exist in an equilibrium state. 

\end{enumerate}

%%%%%%%%%%%%%%%%%%%%%%%%%%%%%%%%%%%%%%%%%%%%%%%%%%%%%%%%%%%%%%%%%%%%%%%%%%%%%%
%%%%%%%%%%%%%%%%%%%%%%%%%%%%%%%%%%%%%%%%%%%%%%%%%%%%%%%%%%%%%%%%%%%%%%%%%%%%%%
%%%%%%%%%%%%%%%%%%%%%%%%%%%%%%%%%%%%%%%%%%%%%%%%%%%%%%%%%%%%%%%%%%%%%%%%%%%%%%
%%%%%%%%%%%%%%%%%%%%%%%%%%%%%%%%%%%%%%%%%%%%%%%%%%%%%%%%%%%%%%%%%%%%%%%%%%%%%%
\acknowledgements
We thank the anonymous referee for his/her valuable comments.
Numerical computations were partially carried out on Aterui
supercomputer at Center for Computational Astrophysics, CfCA, of
National Astronomical Observatory of Japan, and the K computer at the
RIKEN Advanced Institute for Computational Science (Proposal numbers
hp120286 and hp130026).  This work has been funded by MEXT HPCI
STRATEGIC PROGRAM and MEXT/JSPS KAKENHI grant No. 24740115.

\bibliographystyle{apj}

\end{document}